\documentclass[12pt]{article}
\title{Effects of zonal flows on transport crossphase in Dissipative Trapped-Electron Mode turbulence in edge plasmas}
\author{M. Leconte and R. Singh \\
National Fusion Research Institute, Daejeon 34133, South Korea \\
 Email: \textit{mleconte@nfri.re.kr}}
\usepackage{amsmath}
\usepackage{graphicx}

\newcommand{\dif}{\partial}

\newcommand{\ky}{k_y}
\newcommand{\col}{\nu}

\newcommand{\nk}{|n_k|}
\newcommand{\fk}{|\phi_k|}

\newcommand{\freq}{\omega}
\newcommand{\wk}{\omega_k}
\newcommand{\etae}{\eta_e}
\newcommand{\trap}{f_t}

\newcommand{\cpn}{\delta_k}

\newcommand{\nratio}{\beta_k}

\newcommand{\myim}{{\rm Im}}

\renewcommand{\wp}{\omega_0}
\newcommand{\kp}{{k_0}}
\newcommand{\cpnp}{\delta_0}
\newcommand{\fp}{\phi_0}
\newcommand{\np}{n_0}

\newcommand{\wq}{\omega_q}

\newcommand{\fq}{\phi_q}
\newcommand{\nq}{n_q}
\newcommand{\fz}{\phi_z}
\newcommand{\nz}{n_z}
\newcommand{\qx}{q_x}

\newcommand{\fone}{\phi_1}
\newcommand{\none}{n_1}
\newcommand{\cpone}{\delta_1}

\newcommand{\ntwo}{n_2}

\newcommand{\phmm}{\Delta \delta} %phase mismatch

\newcommand{\gd}{\gamma_d}

\begin{document}
\maketitle

\begin{abstract}
Confinement regimes with edge transport barriers occur through the suppression of turbulent (convective) fluxes in the particle and/or thermal channels, i.e. $\Gamma = \sum \sqrt{ |n_k|^2 } \sqrt{ |\phi_k|^2 } \sin \delta_k^{n,\phi}$, \\ and $Q = \sum \sqrt{ |T_k|^2 } \sqrt{ |\phi_k|^2 } \sin \delta_k^{T, \phi}$, respectively, for drift-wave turbulence. The quantity $|\phi_k|^2$ is the turbulence intensity, while $\delta_k^{n,\phi}$ and $\delta_k^{T, \phi}$ are the crossphases.  For H-mode, standard decorrelation theory predicts that it is the turbulence intensity $|\phi_k|^2$ that is mainly affected via flow-induced shearing of turbulent eddies. However, for other regimes (e.g. I-mode, characterized by high energy confinement but low particle confinement), this decrease of turbulence amplitude cannot explain the decoupling of particle v.s. thermal flux, since a suppression of turbulence intensity $|\phi_k|^2$ would necessarily affect both fluxes the same way. Here, we explore a possible new stabilizing mechanism: zonal flows may directly affect the transport crossphase. We show the effect of this novel mechanism on the turbulent particle flux, by using a simple fluid model [Baver et  al., Phys. Plasmas \textbf{9}, 3318 (2002)] for dissipative trapped-electron mode (DTEM), including zonal flows. We first derive the evolution equation for the transport crossphase $\delta_k$  between density and potential fluctuations, including contributions from the $E \times B$ nonlinearity. By using a parametric interaction analysis including the back-reaction on the pump, we obtain a predator-prey like system of equations for the pump amplitude $\phi_p$, the pump crossphase $\delta_p$, the zonal amplitude $\phi_z$ and the triad phase-mismatch $\Delta \delta$. The system displays limit-cycle oscillations where the instantaneous DTEM growth rate - proportional to the crossphase -  shows quasi-periodic relaxations where it departs from that predicted by linear theory. 

%We show that in Dissipative trapped-electron mode (DTEM) turbulence, zonal flows can have a stabilizing effect, not only on the turbulence amplitude, but also directly on the transport crossphase between density and potential. In the framework of the fluid DTEM model [Baver et al., Phys. Plasmas 9, 3318 (2002)], the dynamics of the crossphase is obtained and a parametric interaction analysis is carried out. The result is a predator-prey like model which describes the self-consistent interaction between the amplitude and crossphase of the pump wave and the zonal flow amplitude.
\end{abstract}

\section{Introduction}
Transitions to enhanced confinement regimes such as H-mode play an important role in magnetic fusion devices like ITER.  Such regimes occur through the suppression of the turbulent (convective) fluxes in the particle and/or thermal channel.  For H-mode, theory predicts that it is the amplitude of the turbulence that is mainly affected via flow-induced shearing of turbulent eddies \cite{ConnorWilson2000,DiamondI2Hahm2005, Moyer1995, TynanFujisawaMcKee2009}.
However, for other regimes such as I-mode \cite{Hubbard2016}, characterized by high energy confinement but low particle confinement, this decrease of turbulence amplitude cannot explain the decoupling of particle v.s. thermal flux, since a suppression of amplitude would necessarily affect both fluxes. It is thus important to identify and analyse particle v.s. thermal transport decoupling mechanisms. As convective fluxes depend both on the amplitude but also on the cross-correlation, i.e. crossphase between the two fields, a possible mechanism is the direct modification of the crossphase.
Note that several experiments showed a change in the crossphase between density and potential during the L-H transition \cite{Birkenmeier2013, Kobayashi2017, TynanFujisawaMcKee2009}.
We identify and study such a mechanism for the dissipative trapped-electron mode (DTEM) fluid model \cite{Baver2002, TerryGato2006} in the present work through the effect of zonal flows on the crossphase. The effect of flow-shear on the transport crossphase was investigated in Ref. \cite{Ware1996} for resistive pressure-gradient turbulence. It was shown that the crossphase only depends on the renormalized propagator in this model. Two field models describing trapped-electron mode were analyzed in detail \cite{Gang1991}. The $E \times B$ nonlinearity was found to have an important role via a direct energy cascade to small scale. The crossphase dynamics was analyzed for the Hasegawa-Wakatani model in Ref. \cite{An2017}. It was found that the crossphase associated to the particle flux can be nonlinearly modified compared to the linear theory, thus invalidating the well-known '$i \delta$' approximation . The change resulted in a decrease of the crossphase, i.e. the nonlinearity had a \emph{stabilizing effect}.
We now summarize our main results: i) It is shown that zonal flows can suppress the transport crossphase between density fluctuations $\tilde n$ and potential fluctuations $\tilde \phi$, via the convective $E \times B$ nonlinearity,  ii) This occurs through a nonlinear shift of the crossphase from its linear value,  beyond the standard '$i \delta$' approximation, and iii) This effect originates from the imaginary part of the triplet correlation $\langle \tilde n \tilde \phi \tilde n\rangle$, Eqs. (\ref{cpden},\ref{tripletnonlin1}). The article is organized as follows: In Section $2$, we present the DTEM model, and analyze the associated crossphase dynamics. In Section $3$, we derive the nonlinear 0D model, using a parametric four-wave analysis and present the results. Finally, Section $4$ provides a discussion of the results and conclusions.

\section{Model}

We are interested in the edge region of a fusion device, where collisions are important. Therefore, we consider the regime $\col \gg \omega_k$ where $\col= \nu_{ei} / \epsilon$ is the de-trapping rate for trapped electrons, with $\nu_{ei}$ the electron-ion collision frequency and $\epsilon = a/R$ the inverse aspect ratio . Hence, the fluid approach may be applicable near the plasma edge. We are interested in trapped electron mode transport in this regime, because it is experimentally relevant. We choose the simplest model, to focus on the physics.
We consider the following model of trapped-electron mode (TEM), based on Baver et al. \cite{Baver2002} but modified to include zonal modes, consisting of trapped electron continuity and charge balance:
\begin{eqnarray}
\frac{\dif n}{\dif t} + v_E \cdot \nabla n + (1+ \alpha \etae) \frac{\dif \phi}{\dif y} & = & -\col ( \tilde n - \tilde \phi )
\label{ele00} \\
\frac{\dif}{\dif t} \Big[ (1 - \trap) \tilde \phi -\nabla_\perp^2 \phi \Big] - v_E \cdot \nabla \nabla_\perp^2 \phi
+ \Big[ 1 - \trap (1+ \alpha \etae) \Big] \frac{\dif \phi}{\dif y}
& = & \trap \col ( \tilde n - \tilde \phi)
\label{ion00} \qquad
\end{eqnarray}
with a re-definition of the effective density $n = n_e / \trap + n_{ep} / (1- \trap)$, where $n_e$ is the total density, $n_e = n_{et} + n_{ep}$, $n_{et} = \trap \hat n_e$, $\trap$ the fraction of trapped electrons, and with the fluctuating quantities $\tilde n = n - \langle n \rangle$ and $\tilde \phi = \phi - \langle \phi \rangle$, where the brackets denote a zonal average and $\tilde n_{ep}= (1- \trap)\tilde \phi$ since passing electrons are assumed Boltzmann for non-zonal modes. This modified model reduces to the Charney-Hasegawa-Mima equation in the limit of no trapped-electrons ($\trap \to 0$), thus this model conserves the total potential vorticity $\phi - \nabla_\perp^2 \phi$ in this limit. We thus expect generation of zonal flows in this model since $\trap = \sqrt{\epsilon}$ is small  (typically $\trap \simeq 0.2$) due to the inverse aspect-ratio $\epsilon = a / R \ll 1$. We consider a slab geometry $x,y,z$ for simplicity, though particle trapping is a toroidal phenomenon. This is plausible since we consider high-$n$ modes and high collisions. Here, $x$ is the local radial coordinate, and $y,z$ are the local poloidal and toroidal directions of a fusion device. In the following, we give a schematic derivation of Eqs. (\ref{ele00}, \ref{ion00}). We start from the original model by  Baver et al. [5], valid for non-zonal modes:
\begin{eqnarray}
\frac{\dif n}{\dif t} + v_E \cdot \nabla n +(1+ \alpha \etae) \frac{\dif \phi}{\dif y} & = & - \col (n- \phi) \\
\frac{\dif}{\dif t} [ 1 - \nabla_\perp^2 - \trap ] \phi  -v_E \cdot \nabla \nabla_\perp^2 \phi + [1- \trap (1+ \alpha \etae)] \frac{\dif \phi}{\dif y} & = & \trap \col (n- \phi)
\end{eqnarray}
with $v_E = \hat z \times \nabla \phi$, and the normalizations $(c_s / \rho_s) t \to t$, $\rho_s \nabla_\perp \to \nabla_\perp$. Here
$\trap = n_{et}/n_0$ is the electron trapping fraction, $\etae = L_n / L_{T_e}$ is the ratio of temperature gradient to density gradient, and $\alpha =3/2$. The parameter $\col = \nu_{ei} / \trap$ is the de-trapping rate and characterizes the strength of coupling of the 2 fields. The density $n=n_e/ \trap + \phi$ is an effective electron density, consisting of the passing electron density assumed Boltzmann $(1- \trap) \phi$ and the trapped-electron density $\trap \hat n_e$. In addition, we write the equations for the zonal components, for which electrons are \emph{not Boltzmann}:
\begin{eqnarray}
\frac{\dif n_{zon}}{\dif t} + \frac{\dif}{\dif x} \Big< \tilde v_{Ex} \tilde n_{et} \Big> & = & 0 \\
\frac{\dif}{\dif t} \nabla_\perp^2 \phi_{zon} + \frac{\dif}{\dif x} \Big< \tilde v_{Ex} \nabla_\perp^2 \tilde \phi \Big> & = & 0
\end{eqnarray}
where $n_e = n_{et} + n_{ep}$ with $n_{ep}$ the density of passing electrons, $\tilde n_e = n_e - \langle n_e \rangle$ and
$\tilde \phi = \phi - \langle \phi \rangle$. The zonal density is $n_{zon} = \langle n_{et} \rangle + \langle n_{ep} \rangle = \langle n_{et} \rangle$, that is the zonal density vanishes for passing electrons $\langle n_{ep} \rangle$, since passing electrons don't contribute to the $E \times B$ nonlinearity (the nonlinear drive for zonal density). Here, we used $\tilde n_{ep} = (1-\trap) \tilde \phi$ and thus $\langle \tilde v_{Ex} \tilde n_{ep} \rangle = 0$ since we assume passing electrons to be Boltzmann for non-zonal modes. Combining the equations for non-zonal modes and for zonal modes, we obtain the model (\ref{ele00},\ref{ion00}).

\subsection{Linear analysis}

Linearizing the system of equations, we obtain:
\begin{eqnarray}
-i \freq n_k +i (1+ \alpha \etae) \ky \phi_k & = & \col (\phi_k - n_k)
\label{linele0} \\
-i (1+k_\perp^2 - \trap) \freq \phi_k +i [1 - \trap(1+ \alpha \etae) ] \ky \phi_k & = & - \trap \col (\phi_k - n_k)
\label{linion0}
\end{eqnarray}
with $k_\perp^2 = k_x^2 + \ky^2$ the squared perpendicular wavenumber.

In matrix form:
\begin{equation}
\begin{bmatrix}
-i \freq + \col & (1+ \alpha \etae) i \ky -\col \\
- \trap \col & -i (1+k_\perp^2 - \trap) \freq + i [ 1- \trap(1+\alpha \etae) ] \ky + \trap \col 
\end{bmatrix}
\begin{bmatrix}
n_k \\
\phi_k
\end{bmatrix}
= 0
\end{equation}

The associated linear dispersion relation is given by:
\begin{align}
\Big[ -i \freq + \col \Big] \Big[  -i (1+k_\perp^2 - \trap) \freq + i [ 1- \trap(1+\alpha \etae) ] \ky + \trap \col \Big] 
- \trap \col \Big[ \col - (1+ \alpha \etae ) i \ky \Big] = 0.
\end{align}

After a little algebra, this reduces to:
\begin{equation}
(1+k_\perp^2 - \trap) \freq^2 + i \col \Big( 1+k_\perp^2 + \frac{i}{\col} [ 1- \trap(1+\alpha \etae) ] \ky   \Big) \freq
- i \ky \col = 0.
\label{disprel0}
\end{equation}

% ion continuity analysis

A more physics-based linear analysis can be obtained by evaluating first the density response.
One can obtain the effective density response from Eq. (\ref{ele00}) as:
\begin{equation}
n_k \simeq \left[ 1 - \frac{i [(1+ \alpha \etae) \ky - \freq] }{\col} \right] \phi_k, 
\end{equation}
in the limit $|\omega| \ll \col$. This could be interpreted as an $'i \delta'$ prescription: $n_k = (1 - i \delta) \phi_k$, with $\delta  = [(1+ \alpha\etae) \ky - \freq ] / \col$.

Multiplying Eq. (\ref{linele0}) by $\trap$, adding to Eq. (\ref{linion0}) and replacing the electron response in the resulting equation, the approximate linear dispersion relation takes the form:
\begin{equation}
(1+k_\perp^2) \freq - \ky = i \trap \delta \freq
\label{disprel1}
\end{equation}
%which is identical to (\ref{disprel0}), but makes it easier to identify the physics, and derive limiting expressions.
In the relevant dissipative regime $\omega \ll \col$ the dispersion relation can be solved by a perturbation in the small parameter $(\omega / \col) \ll 1$. We expand the complex frequency as: $\omega = \omega^{(0)} + \omega^{(1)}$. At zeroth-order, Eq. (\ref{disprel1}) yields the drift-wave frequency $\omega_R = \ky / (1+k_\perp^2)$.
At first order, Eq. (\ref{disprel1})  yields:
\begin{equation}
\gamma(k_x, \ky) \simeq  \frac{\trap}{\col} \left[  \frac{\alpha \etae \ky^2}{(1+k_\perp^2)^2
} + k_\perp^2 \frac{ \ky^2 }{(1+ k_\perp^2)^3}  \right] 
\label{growth}
\end{equation}
where $\gamma = {\rm Im} ~ \omega$ denotes the linear growth-rate, and we replaced $\delta$ by its expression.

\begin{figure}
\begin{center}
\includegraphics[width=0.5\linewidth]{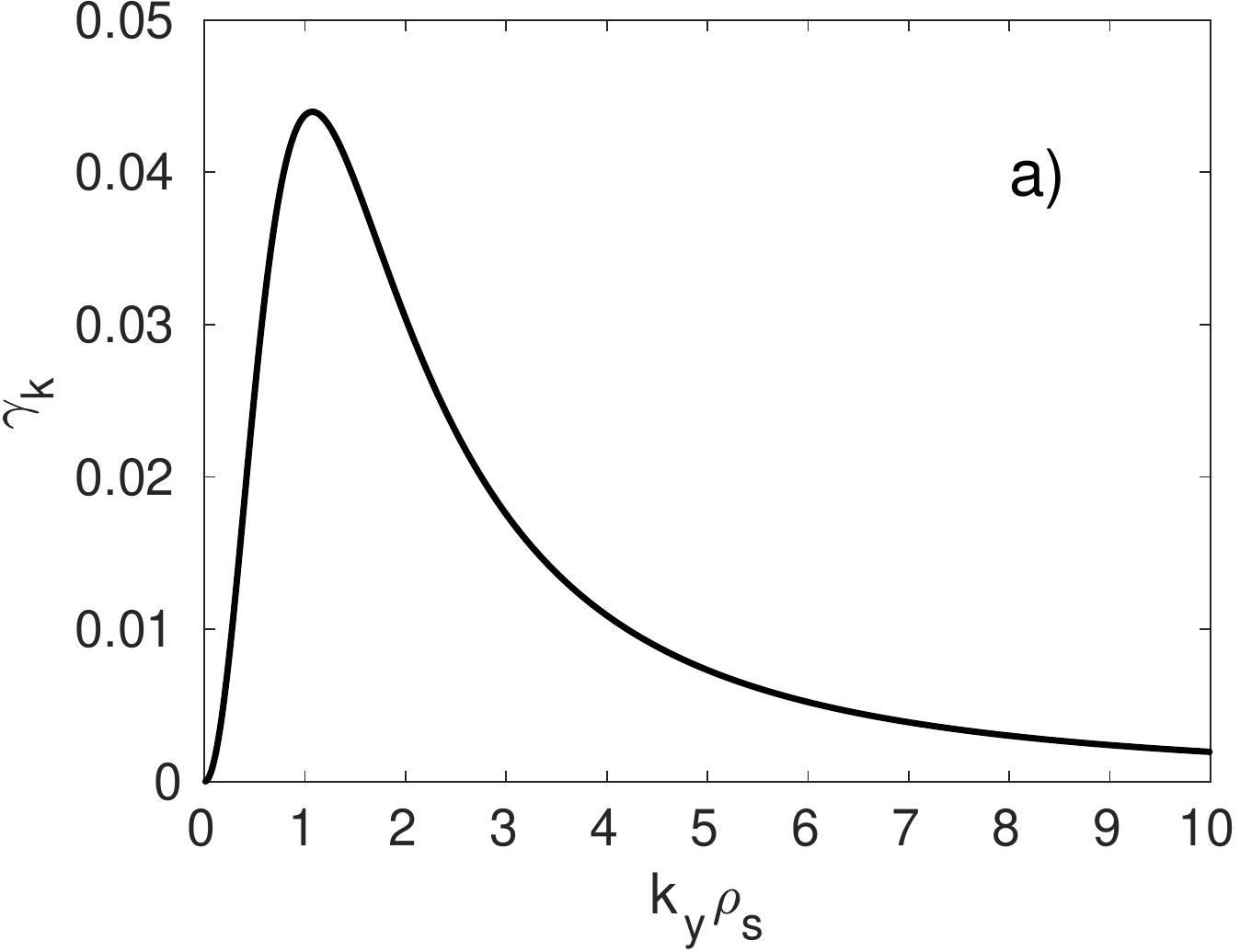}\includegraphics[width=0.5\linewidth]{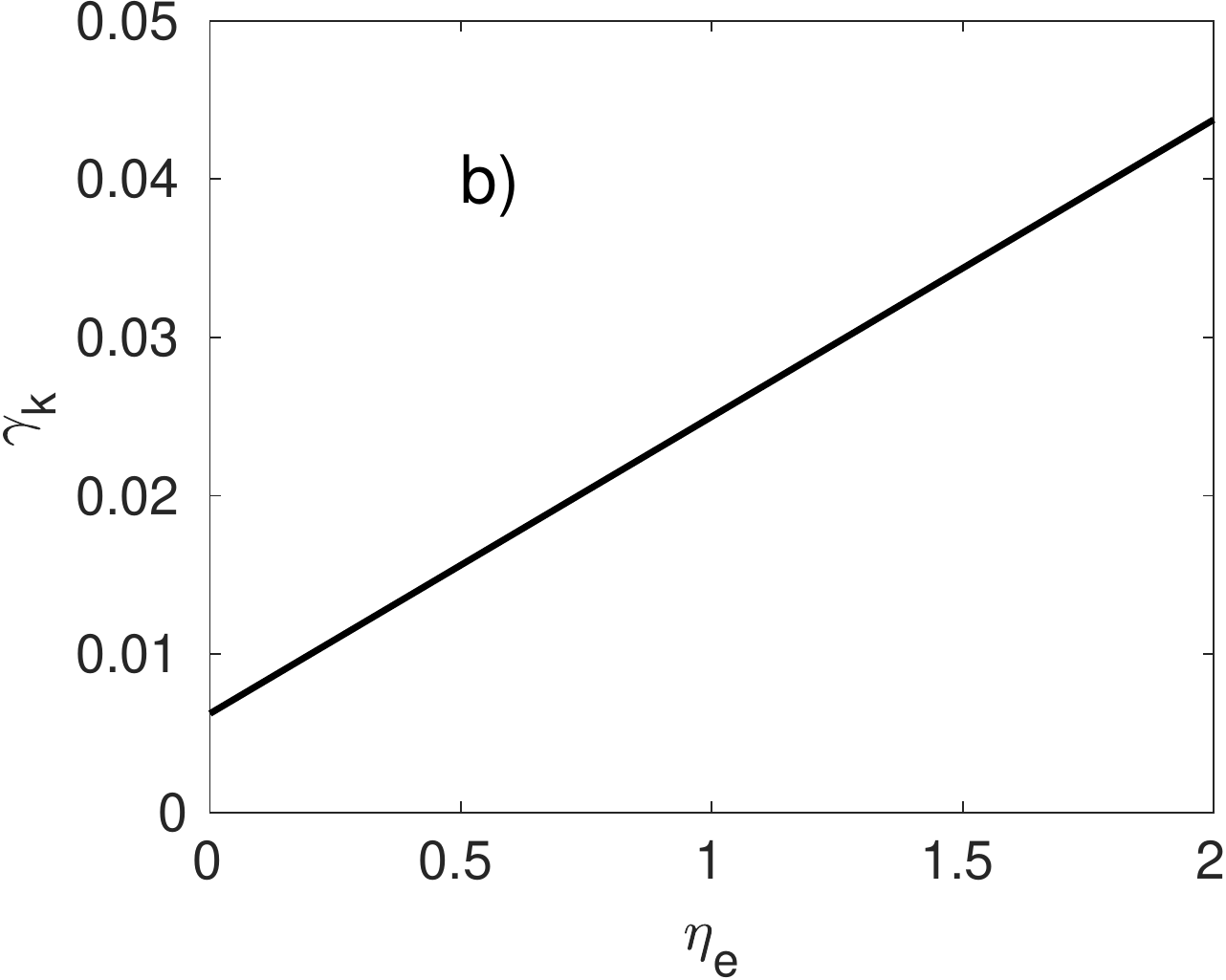}
\caption{(a) Linear growth-rate v.s. wavenumber $k_y$ for the parameters $\col=10$ and $\etae=2$. The growth-rate is given by Eq. (\ref{growth}). (b): linear growth-rate v.s. stablity parameter $\eta_e$, for $\ky \rho_s = 1$.}
\label{growthrate}
\end{center}
\end{figure}

The linear growth-rate $\gamma$, Eq. (\ref{growth}) is shown v.s. wavenumber $k_y$, for $\etae = 0.1$ and $\etae = 2$ [Fig. \ref{growthrate}]. The linear growth-rate shows a peak around $\ky \rho_s \le 1$. The instability does not show a threshold behavior, unless CTEMs. This is a short-coming of the simplified model, which does not evolve the trapped  electron temperature.

%The full dispersion relation (\ref{disprel0}) or (\ref{disprel1}) is best solved using a numerical method, e.g. the Nelder-Mead function minimization technique \cite{NelderMead1965}. We show the results, using this method, for two values of the  adiabaticity parameter $C=5,10$, for  $\pardiff=1$ and $\etae = 0.2$ [Fig \ref{growthrate}].

% CROSSPHASE DYNAMICS

\subsection{Cross-phase dynamics}

Considering a Fourier representation of a physical quantity, $g = \sum_k g_k e^{i {\bf k} \cdot {\bf r}} e^{-i \omega_k t}$, with $g=n, \phi$, we can write the complex amplitudes $n_k,\phi_k$ in amplitude-phase form \cite{Ritz1989}:
\begin{eqnarray}
n_k & = & \nk  \exp (-i \cpn)  \\
\phi_k& = & \fk
\end{eqnarray}
where $\cpn$ denotes the 'density crossphase', defined as the phase-angle of density \emph{with respect to the potential fluctuation with the same mode number}. Note the minus sign in the definition of the crossphase. This convention gives a {\emph positive} crossphase during phase-locking and is thus convenient to compare with the well-known ''$i \delta$'' approximation.

Using this ansatz, the electron continuity Eq. (\ref{ele00}) becomes, for non-zonal modes with ${\bf k} \neq {\bf q}$:
\begin{align}
e^{-i \cpn} \frac{\dif \nk }{ \dif t} +i e^{-i \cpn} \nk \left[ - \frac{\dif \cpn}{ \dif t} - \wk \right]
=
- i (1+ \alpha \etae) \ky \fk
+\col \Big[ \fk - \nk e^{-i \cpn} \Big] \notag\\
- \tilde v_E \cdot \nabla \tilde n,
\label{cont1}
\end{align}

\noindent and for zonal modes with ${\bf k} = {\bf q}$:
\begin{align}
e^{-i \cpn} \frac{\dif \nk }{ \dif t} -i e^{-i \cpn} \nk \frac{\dif \cpn}{ \dif t}
= - \Big< \tilde v_E \cdot \nabla \tilde n \Big>,
\label{cont1-zon}
\end{align}

We separate the real and imaginary parts and derive the equations for the amplitudes (real part) and phase-angles (imaginary part). From the imaginary part, we obtain after some algebra, for non-zonal modes:
\begin{align}
\nk  \left[  -\frac{\dif \cpn}{\dif t} - \wk \right] & = & -(1+ \alpha \etae)\ky \fk \cos\cpn
- \col \fk \sin \cpn 
-  {\rm Im} \{ e^{i \cpn} \tilde v_E \cdot \nabla \tilde n \}
\label{nphase}
\end{align}
which reduces to:
\begin{equation}
\frac{\dif \cpn}{\dif t} = (1+ \alpha \etae) \ky \nratio \cos \cpn - \wk - \col \nratio \sin \cpn
+N_k,
\label{cpden}
\end{equation}
where
$\nratio = \fk / \nk$ denotes the amplitude ratio and
 $N_k$ is  the nonlinear contribution, due to the \emph{ $E \times B$ nonlinearity}, given by:
\begin{equation}
N_k = \frac{1}{\nk^2} {\rm Im} \{ n_k^* ~ (v_E \cdot \nabla n)_k \},
\label{tripletnonlin1}
\end{equation}
where convolution in wavenumber space is implicit, and denoted by $(\ldots)_k$.
We recover similar results as Ref. \cite{An2017} for the dynamics of the crossphase  $\cpn$ (Eq. 5 in Ref \cite{An2017}).

For zonal modes, one obtains:
\begin{equation}
\frac{\dif \delta_q}{\dif t} = \frac{1}{|n_q|^2} {\rm Im} \left\{ n_q^* ~  \Big< v_E \cdot \nabla n \Big> \right\}
\label{cpden-zon}
\end{equation}

For the linear analysis of cross-phase dynamics, Eq. (\ref{cpden}) - with $N_k=0$ - is self-consistent, assuming the amplitude ratio is given. \\
However, for the \emph{nonlinear dynamics}, the term $N_k$ - which involves the resonant triads
$({\bf k},{\bf k}', {\bf k} + {\bf k}')$ and $({\bf k},{\bf k}',{\bf k}-{\bf k}')$ - must be evaluated.

To express the amplitude ratio, we write the real part of Eq. (\ref{cont1}):
\begin{equation}
\frac{1}{\nk} \frac{\dif \nk}{\dif t} = (1+ \alpha \etae) \ky \nratio \sin \cpn +\col \Big( \nratio \cos \cpn - 1 \Big)
- \frac{1}{\nk^2} {\rm Re} \Big\{ n_k^* (v_E \cdot \nabla n)_k \Big\}
\label{ele1}
\end{equation}

Eq. (\ref{ele1}) can be used to find a relation between the amplitude ratio $\nratio$ and the crossphase $\cpn$.
Neglecting the quadratic nonlinearity, one obtains:
\begin{equation}
\frac{1}{\nk} \frac{\dif \nk}{\dif t} \simeq
(1+ \alpha \etae) \ky \nratio \sin \cpn
+ \col \Big[\nratio \cos \cpn -1 \Big]
\label{ampratio0}
\end{equation}
with $\nratio = \fk / \nk$. Using the approximation of small growth-rate $|\gamma_k| \ll |\omega_k|$, i.e. $\dif_t \nk / \nk \simeq 0$, we obtain the algebraic equation:
\begin{equation}
\nratio \left[ 1 + \frac{(1+ \alpha \etae) \ky}{\col} \tan \cpn \right] \cos \cpn - 1 \simeq 0
\label{ampratio1}
\end{equation}

To lowest order, this gives, for linearly unstable modes:
\begin{equation}
\nratio \simeq \left[ 1- \frac{(1+ \alpha \etae) \ky}{\col} \tan \cpn \right] \frac{1}{\cos \cpn}
\end{equation}

\begin{figure}
\begin{center}
\includegraphics[width=0.5\linewidth]{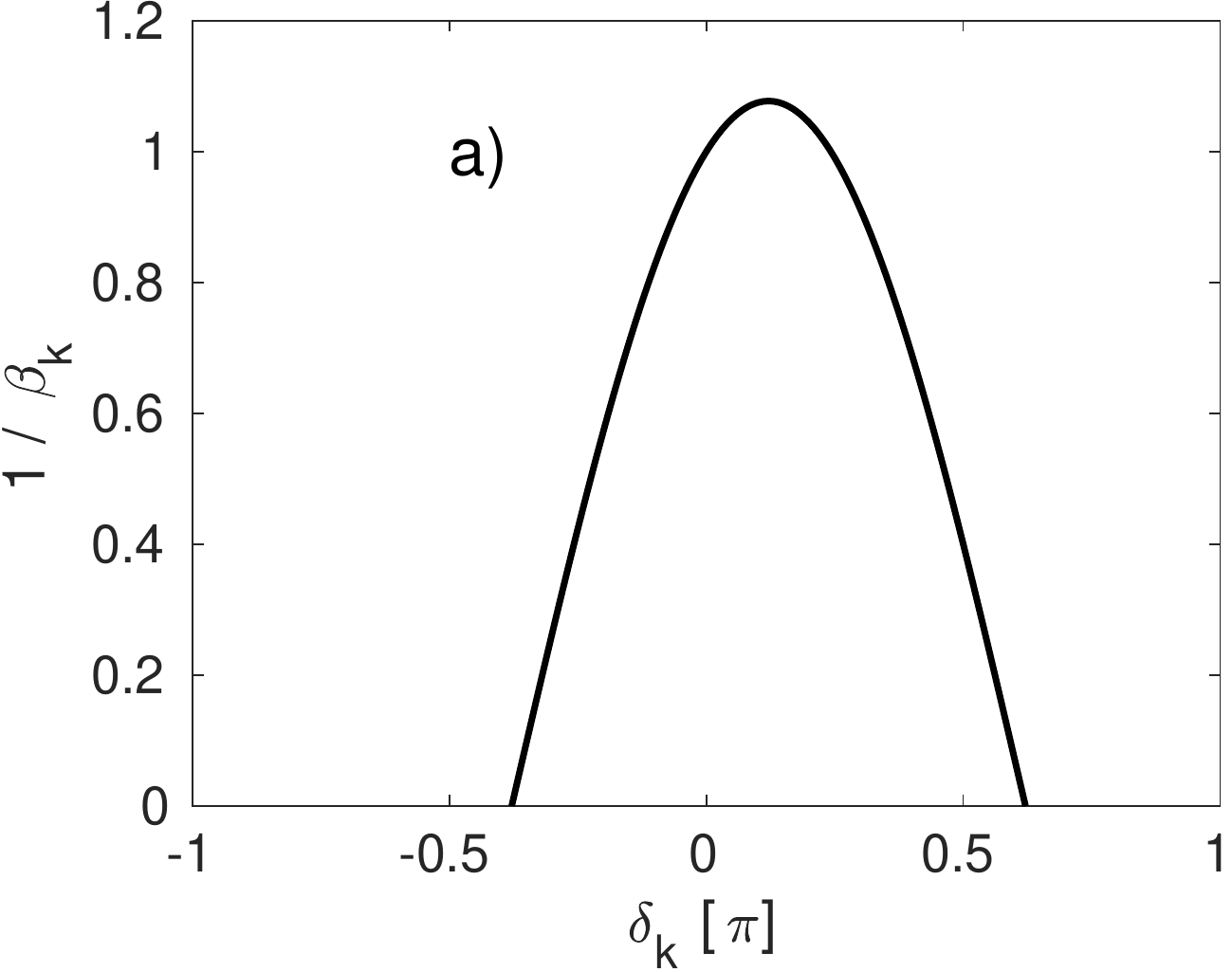}\includegraphics[width=0.5\linewidth]{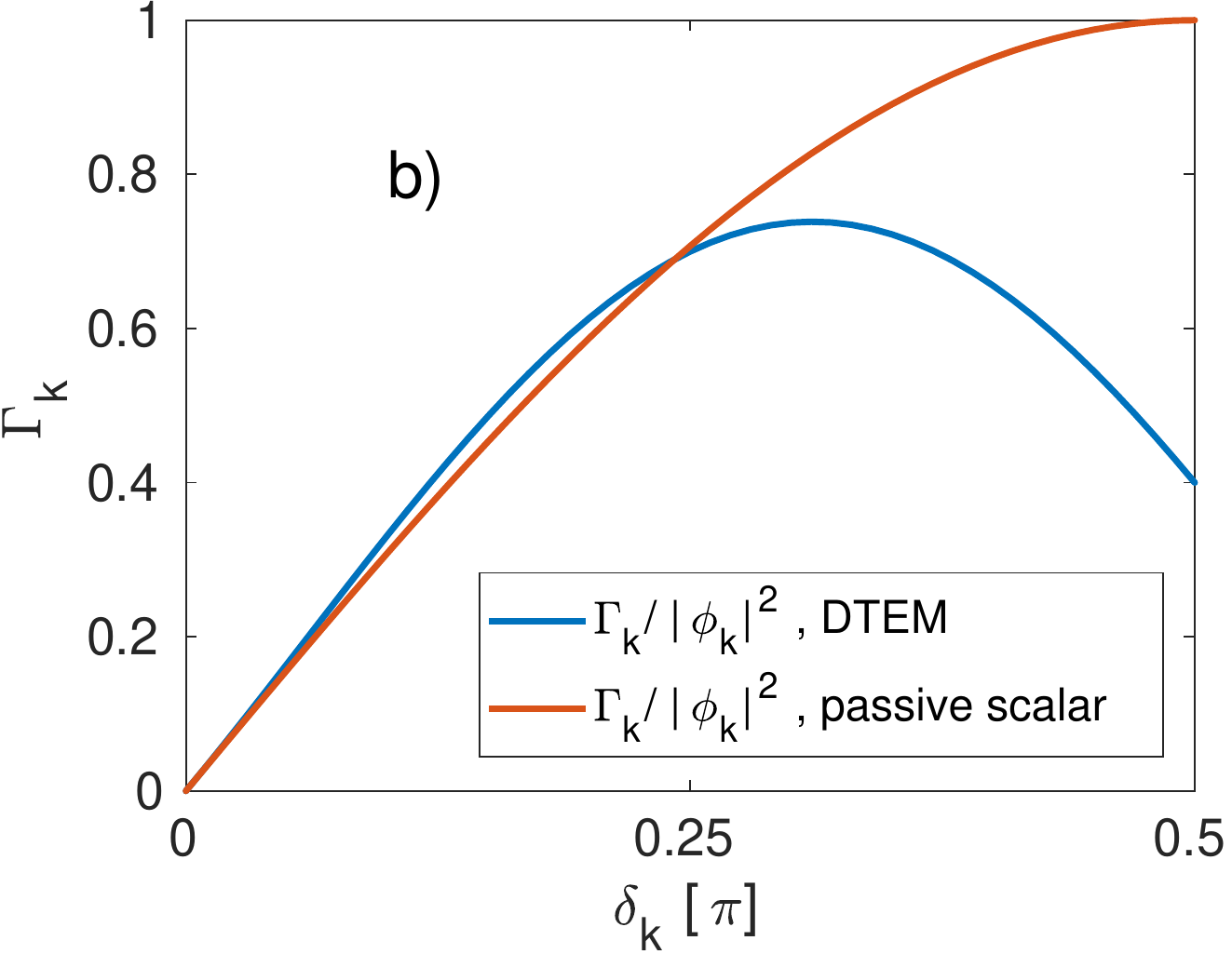}
\caption{ (a) Inverse of amplitude ratio $1 / \nratio$ v.s. crossphase $\cpn$, for $\nu=10$, $\etae = 2$, at $k_y \rho_s = 1$, (b) The ratio of particle flux with the wavenumber $k$,  $\Gamma_k$ to turbulence square amplitude $\fk^2$, for DTEM (red) and for a passive scalar model (blue).}
\label{fig-ampratio}
\end{center}
\end{figure}

The amplitude ratio $\nratio$ is shown v.s. the crossphase, for linearly-unstable modes  [Fig.  \ref{fig-ampratio}a]. Since for DTEM, the crossphase is small $|\cpn| \ll 1$, Fig. \ref{fig-ampratio}a shows that the amplitude ratio is $\nratio \simeq 1$ at lowest-order. Fig \ref{fig-ampratio}b shows that the ratio of particle flux to turbulence square amplitude $\Gamma_k / \fk^2$ is maximal at the crossphase $\delta_k \simeq \pi / 4$, contrary to a simple passive scalar model, where it would peak at $\delta_k = \pi / 2$. This difference is due to the amplitude ratio dependence on the crossphase in DTEM.

Note that the dynamical evolution of the crossphase implies the dynamical change in particle flux. Therefore, we can derive the following equation describing the \emph{relaxation dynamics of the particle flux} with wavenumber ${\bf k}$:
\begin{align}
\left[ \frac{\dif }{\dif t} + \left( 1 + \frac{\trap}{1 + k_\perp^2 - \trap} \right) \col \right] \Gamma_k
- \trap (1+ \alpha \etae ) \ky^2 \fk^2
+  \frac{ [1- \trap (1+\alpha \etae) ] \ky^2}{1 + k_\perp^2 - \trap} {\rm Re} (n_k^* \phi_k)
= \notag\\
 - \trap \ky \cdot \frac{1}{2} \sum_{k = k' + k''} (\hat z \times k') \cdot k''  {\rm Im} ( \phi_k n_{k'}^* \phi_{k''}^* - \phi_k \phi_{k'}^* n_{k''}^*) \notag\\
+ \frac{\trap \ky}{1 + k_\perp^2 - \trap} \sum_{k = k' + k''} (k_\perp'^2 - k_\perp''^2) (\hat z \times k') \cdot k''
 {\rm Im} ( n_k^* \phi_{k'} \phi_{k''})
\label{fluxevo0}
\end{align}\
The derivation is given in Appendix.
The particle flux with the wavenumber ${\bf k}$ is defined as $\Gamma_k = \ky ~\myim \{ n_k^* \phi_k \}$. A similar equation was derived in Ref. \cite{Gang1991} for the complex-valued cross-correlation $\langle \tilde n \tilde \phi \rangle$ of the Hasegawa-Wakatani model.
Equation (\ref{fluxevo0}) clearly shows that the flux does not respond instantaneously to the driving gradient $(1+ \alpha \etae)$, and it is not directly proportional to the gradient, contrary to a simple Fick's law of diffusion. Instead, there is a finite \emph{response time} or relaxation time $\tau$. We expect this feature to be generic, but the expression for the relaxation time is model-specific. For the simple fluid DTEM model being used in our article, it scales as: $\tau \sim \col^{-1}$, with $\col$ the de-trapping rate.
Noting the similarity of Eq. (\ref{fluxevo0}) with the crossphase dynamics Eq. (\ref{cpden}), we infer that crossphase dynamics mirrors the relaxation dynamics of the particle flux. In fact, writing explicitely the diamagnetic drift dependence on the gradient $v_* \propto - \dif_x \langle n \rangle$, we have the following two coupled equations for crossphase and density gradient:
\begin{eqnarray}
\tau \frac{\dif \cpn}{\dif t} & = &
 - \Big[ \frac{(1+ \alpha \etae) \ky}{\col} -\frac{\ky}{(1+ \ky^2) \col} \Big] \frac{\dif \langle n \rangle}{\dif x} 
- \cpn
 + \frac{1}{\col} N_k \\
\frac{\dif \langle n \rangle}{\dif t} & =  & - \frac{\dif}{\dif x} \left[ \sum \trap \ky \fk^2 \delta_k \right]
+ D_\perp^{res} \frac{\dif^2 \langle n \rangle}{\dif x^2}
+S,
\end{eqnarray}
for $|\delta_k| \ll 1$, with $\tau = 1/ \col$.
Only in the limit of infinite de-trapping rate $\col \to + \infty$, i.e. $\tau \to 0$ and no nonlinear effects $N_k \to 0$ do we recover a local diffusive behaviour $\delta_k \propto - \frac{\dif \langle n \rangle}{\dif x}$.

% PARAMETRIC INTERACTION

\section{Parametric analysis}
It is apparent from Eq. (\ref{cpden}) that wave-wave interactions are responsible for the nonlinear dynamics of 'density crossphase' $\cpn$. Because zonal flows are not linearly unstable and are thus not affected by the linear damping mechanisms (trapped-passing collisons for DTEM) that affect linearly unstable modes, they play a dominant role in wave-wave interactions.
To obtain some physical insight, we consider a simpler setting based on the parametric interaction between TEM drift-waves and zonal flows, i.e. a 4 wave interaction \cite{LashmoreDavies2001, Goswami2000, Singh2001, ChenLinWhite2000, Singh2017, LeconteSingh2018}. A pump drift-wave at $(\wp, \kp)$ interacts with a seed zonal flow at $(\wq, {\bf q} = q_x \hat x)$ to generate two sidebands at $(\omega_{1,2}, {\bf k}_{1,2})$, with the triad resonance condition $\omega_{1,2} = \wp \pm \wq$ and ${\bf k}_{1,2} = {\bf k}_0 \pm {\bf q}$. In turn, the sidebands interact with the pump to nonlinearly drive the zonal flow. Due to energy conservation, we also consider the {\emph{back-reaction} on the 'pump': the zonal flow interacts with the sidebands to back-react on the 'pump.' The parametric interaction is shown on a schematic diagram [Fig. \ref{fig-paraminter}].  \\

\begin{figure}
\begin{center}
\includegraphics[width=0.5\linewidth]{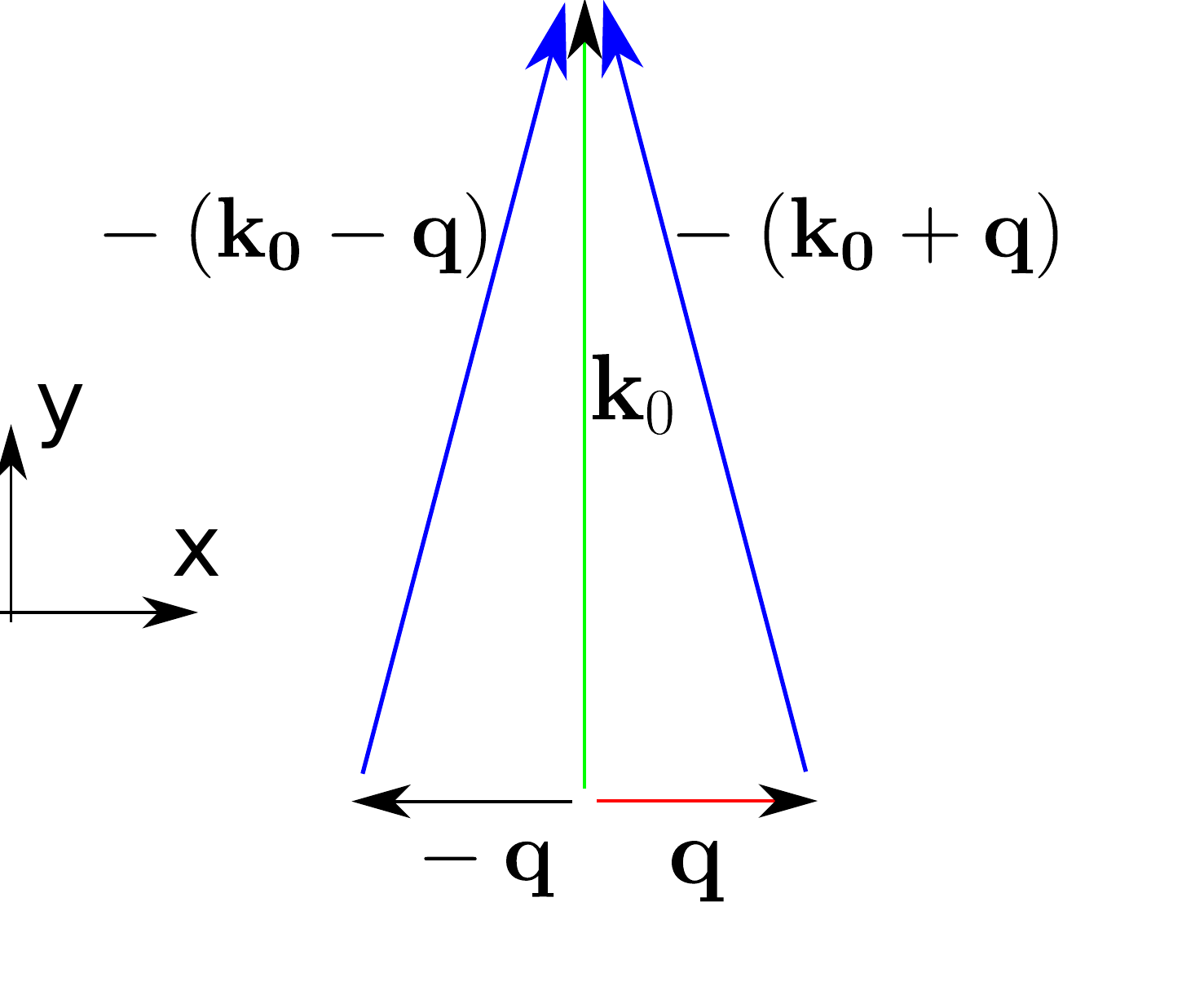}\includegraphics[width=0.5\linewidth]{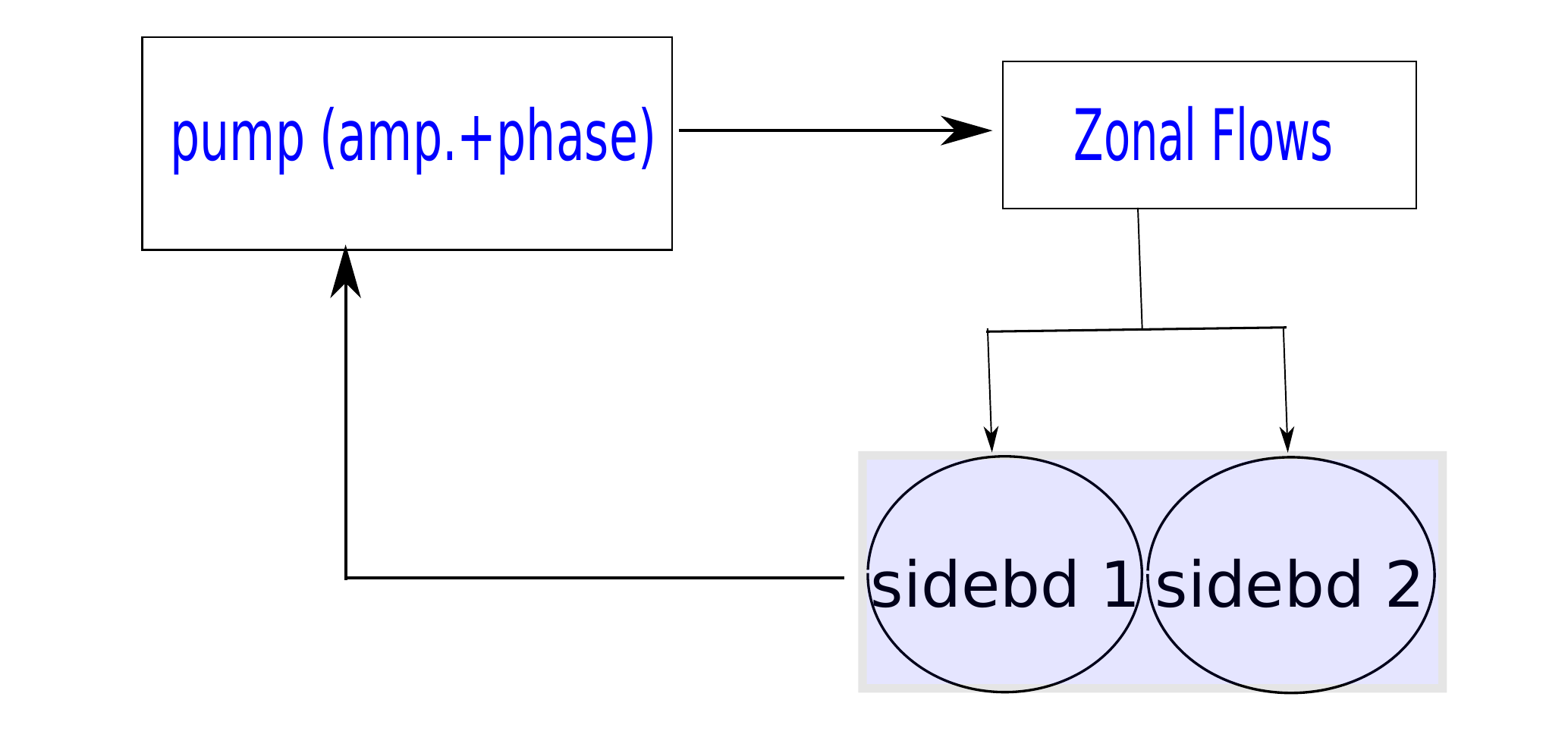}
\caption{A schematic diagram representing the parametric interaction among the pump wave (${\bf k}_0, \wp$), zonal mode (${\bf q}, \omega_q$) and two sidebands.}
\label{fig-paraminter}
\end{center}
\end{figure}

The pump wave is taken as:
\begin{equation}
\begin{bmatrix}
n \\
\phi
\end{bmatrix}
=
\begin{bmatrix}
n_{k0} \\
\phi_{k0}
\end{bmatrix}
\exp [i {\bf k}_0 \cdot {\bf r} - i \wp t] +c.c.,
\end{equation}
with $\wp$ the drift-wave frequency.
The zonal flow is taken in the form:
\begin{equation}
V_{zon} = i \qx \fq \exp(i \qx x - i \wq t) +c.c.,
\end{equation}
and same for zonal density ($n_q$).
The sidebands are written as:
\begin{equation}
\begin{bmatrix}
n_{1,2} \\
\phi_{1,2}
\end{bmatrix}
=
\begin{bmatrix}
\tilde n_{1,2} \\
\tilde \phi_{1,2}
\end{bmatrix}
\exp [i ({\bf k}_0 \pm {\bf q}) \cdot {\bf r} - i \omega_{1,2} t] +c.c.,
\end{equation}
We use a decomposition into the amplitude and the phase, and allow a finite phase-shift for the density, corresponding to the crossphase:
\begin{equation}
\begin{bmatrix}
n_{k0}  \\
\phi_{k0}
\end{bmatrix}
=
\begin{bmatrix}
\np \exp(-i \cpnp) \\
\fp
\end{bmatrix} ,
\end{equation}
and for zonal flows, $\fq = \fz$. \\

% PARAMETRIC INTERACTION RESULT

In the framework of the parametric interaction analysis, one can analytically evaluate the nonlinear term in Eq. (\ref{cpden}).The crossphase dynamics Eq.  (\ref{cpden})  reduces to:
\begin{eqnarray}
\frac{\dif \cpnp}{\dif t} & = &
- \wp  + (1+ \alpha \etae) \kp \beta_0 \cos \cpnp
- \col \beta_0 \sin \cpnp
 +N_0,
\label{cpden1}
\end{eqnarray}

\noindent where $N_0$ is the nonlinear crossphase shift associated to the $E \times B$ nonlinearity, and takes the form:
\begin{equation}
N_0 = - \qx \kp \Big[ \frac{\fq  \none}{\np} \sin (\delta_1 - \cpnp) + \frac{\nq \fone}{\np} \sin \cpnp
+\frac{\fq  \ntwo}{\np} \sin (\delta_2 -\cpnp) + \frac{\nq \phi_2}{\np} \sin \cpnp \Big].
\end{equation}

Using the definition of the amplitude ratio $\beta_k = \fk / \nk$, the nonlinear crossphase shift can be rewritten as:
\begin{equation}
N_0 = - \qx \kp \beta_0 \Big[ \frac{\fz \none}{\fp} \sin (\delta_1 - \delta_0) + \frac{\nz \fone}{\fp} \sin \cpnp
+ \frac{\fz  n_2}{\fp} \sin (\delta_2 - \delta_0) + \frac{\nz \phi_2}{\fp} \sin \cpnp  \Big],
\end{equation}
where the pump amplitude ratio $\beta_0=\fp/\np$ is slaved to the crossphase:
\begin{eqnarray}
\beta_0 & \simeq & \left[ 1 - \frac{(1+ \alpha \etae) \ky}{\col} \tan \cpnp \right] \frac{1}{ \cos \cpnp},
\end{eqnarray}
%whereas the sideband amplitude ratios depend on the sideband crossphases \emph{and} on the $E \times B$ nonlinearity:
%\begin{eqnarray}
%\beta_1 & = & \beta_1(\delta_1, N_1) \\
%\beta_2 & = & \beta_2(\delta_2, N_2)
%\end{eqnarray}

\subsection*{Derivation of the dynamics of zonal modes}

Here we derive the equation for zonal potential (zonal flows) and zonal density. We start from the conservation of potential vorticity:
\begin{equation}
\frac{\dif}{\dif t} \Big[ n_i - \nabla_\perp^2 \phi \Big]
+ {\bf v}_E \cdot \nabla ( n_i- \nabla_\perp^2 \phi) + \frac{\dif \phi}{\dif y} = 0
\end{equation}
with $n_i - \nabla_\perp^2 \phi$ the potential vorticity, $n_i$ the ion density and $n_i = n_{ep} + n_{et}$ due to quasineutrality, with $n_{et} = \trap \hat n_e$.
Flux-surface averaging yields the zonal potential vorticity evolution:
\begin{equation}
\frac{\dif}{\dif t} \left[ n_{zon} - \frac{\dif^2  \phi_{zon} }{\dif x^2} \right]
+ \frac{\dif}{\dif x} \Big[ \Big<  \tilde v_{Ex} \tilde n_{et} \Big> - \Big< \tilde v_{Ex} \nabla_\perp^2 \tilde \phi \Big> \Big] = 0
\label{ap-pvzon}
\end{equation}
with $n_{zon} = \langle n_{et} \rangle$, since passing electrons do not contribute to zonal density.
In addition, the dynamics of effective electron density Eq. (\ref{ele00}) is:
\begin{equation}
\frac{\dif n}{\dif t} + v_E \cdot \nabla n + (1+ \alpha \etae) \frac{\dif \phi}{\dif y} = -\col ( \tilde n - \tilde \phi)
\end{equation}
Flux-surface averaging yields:
\begin{equation}
\frac{\dif n_{zon}}{\dif t} + \frac{\dif}{\dif x} \Big< \tilde v_{Ex} \tilde n_{et} \Big> = 0
\end{equation}
Combining with Eq. (\ref{ap-pvzon}) yields the dynamics of zonal potential:
\begin{equation}
\frac{\dif^2}{\dif x^2} \frac{\dif \phi_{zon}}{\dif t} + \frac{\dif}{\dif x} \Big< \tilde v_{Ex} \nabla_\perp^2 \tilde \phi \Big> = 0
\end{equation}

In Fourier space, the two equations for zonal modes become:
\begin{eqnarray}
\frac{\dif n_q}{\dif t} 	& = &  \sum_k (\hat z \times {\bf q}) \cdot {\bf k} \frac{1}{2} \Big( n_k^* \phi_{k+q} - \phi_k^* n_{k+q} \Big) \\
\qx^2 \frac{\dif \phi_q}{\dif t} 	& = & \sum_k (\hat z \times {\bf q}) \cdot {\bf k} ~( |{\bf k}+{\bf q}|^2-k^2) \phi_k^* \phi_{k+q}
\end{eqnarray}

Using the parametric interaction analysis, the equations for zonal modes reduce to:
\begin{eqnarray}
\frac{\dif \nq}{\dif t}	& = &  \frac{1}{2} \Lambda  \Big[ n_{\kp}^* \phi_{k_1} - \phi_{\kp}^* n_{k_1} \Big] 
+ \frac{1}{2}  \Lambda \Big[ n_{\kp} \phi_{k_2}^* - \phi_{\kp} n_{k_2}^* \Big] \\
\frac{\dif \fq}{\dif t}	& = & \Lambda \Big[ \phi_\kp^* \phi_{k_1} + \phi_\kp \phi_{k_2}^* \Big]
\end{eqnarray}
with ${\bf k}_{1,2} = \kp {\bf \hat y} \pm \qx {\bf \hat x}$ and $\Lambda = \qx \kp$.

% ODE SYSTEM

Using the symmetry of sidebands, we obtain , for $|\cpnp| \ll 1$, the following system of coupled equations:
\begin{eqnarray}
\frac{\dif \cpnp}{\dif t} & = & (1+\alpha \etae)\kp  - \wp  - \col \cpnp + \Lambda \left[ \frac{\fz \none}{\fp} \phmm - \frac{\nz \fone}{\fp} \cpnp \right]
\label{ph-0d} \\
(1+ \kp^2) \frac{\dif \fp}{\dif t} & = & \trap (1+ \alpha \etae) \kp \fp \cpnp - 2 \kp^2 \Lambda \fone \fz
- \trap \Lambda ( \fz \none \cos \cpone - \nz \fone )
\label{pump-0d} \\
\frac{\dif \fz}{\dif t} & = & 2 \Lambda \fp \fone - \mu \fz
\label{zonal-0d} \\
\frac{\dif \nz}{\dif t} & = & \Lambda (\fp \none \cos \cpone - \fp \fone)
\label{zonaln-0d} \\
(1+ k_1^2 - \trap) \frac{\dif \fone}{\dif t} & = & \trap \col (\none \cos \cpone   -\fone) +(\kp^2-\qx^2) \Lambda \fp \fz
\label{sidb-0d} \\
\frac{\dif \none}{\dif t} & = & - \col (\none - \fone \cos \cpone) - \frac{\Lambda}{2} ( \fp \nz - \fp \fz )
\label{sidbn-0d} \\
\frac{\dif \phmm}{\dif t} & = & - \Delta \omega - \col \phmm  \notag\\
 & & - \frac{\Lambda}{2} \left[ \left( \frac{ \fp \fz - \fp \nz }{\none} - 2 \frac{ \fz \none}{\fp} \right) \phmm
+ \Big( \frac{\fp \nz}{\none} + 2 \frac{\nz \fone}{\fp} \Big) \cpnp \right]
\label{mismatch-0d} \qquad
\end{eqnarray}
with $\phmm = \cpnp - \cpone$, and $\cos \cpone \simeq 1 - \cpone^2/2$. The derivation is given in Appendix

We now give a description of this system of equations. Eq. (\ref{ph-0d}) describes the dynamics of the crossphase $\cpnp$ associated to the pump. It takes the form of a Kuramoto-like equation, a paradigm for describing synchronization phenomena in populations of coupled oscillators \cite{Acebron2005}. The first and second terms on the r.h.s. represent the difference between the pump frequency $\wp$ and the electron diamagnetic frequency $(1+ \alpha \etae) \kp$, including the $\nabla T_e$ contribution. The third term on the r.h.s. is the analog of the 'pinning term' of the Kuramoto equation responsible for phase-locking, here due to collisions between trapped and passing electrons. The last term on the r.h.s. comes from the $E \times B$ nonlinearity and involves both zonal flows ($\fz$) and zonal density ($\nz$).
Eq. (\ref{pump-0d}) is the evolution equation for the pump amplitude $\fp$. The first term on the r.h.s. is the linear drive, which crucially depends on the pump crossphase. The second term on the r.h.s. is the usual zonal flow shearing effect (always stabilizing) due to the polarization nonlinearity. The last term on the r.h.s. is due to the $E \times B$ nonlinearity, and involves both zonal flows ($\fz$) and zonal density ($\nz$).
Eqs. (\ref{zonal-0d}) and (\ref{zonaln-0d}) describe the dynamics of zonal modes. Zonal flows ($\fz$) are nonlinearly driven due to the polarization nonlinearity and linearly damped due to neoclassical friction (respectively 1st and 2nd term on r.h.s. of (\ref{zonal-0d})).
Zonal density ($\nz$) is nonlinearly driven due to the $E \times B$ nonlinearity, r.h.s. of Eq. (\ref{zonaln-0d}).
Eqs. (\ref{sidb-0d}) and (\ref{sidbn-0d}) describe the dynamics of the sideband potential ($\fone$) and density ($\none$), respectively. These are linearly coupled via collisions ($\col$). The sideband potential is nonlinearly driven via the polarization nonlinearity, last term on the r.h.s. of Eq. (\ref{sidb-0d}), while the sideband density is nonlinearly driven via the $E \times B$ nonlinearity.
Finally, Eq. (\ref{mismatch-0d}) describes the dynamics of \emph{phase-mismatch} $\Delta \delta = \cpnp - \delta_1$. The first term on the r.hs. is the negative of frequency mismatch $\Delta \omega = \wp - \omega_1$, and the 2nd term on the r.h.s. is due to collisions. The two last terms on the r.h.s. are due to the $E \times B$ nonlinearity. \\

This 0D model is an extension of that presented in Ref. \cite{ChenLinWhite2000}, to self-consistently include the crossphase dynamics, with a novel stabilizing effect of zonal flows on the transport crossphase.
From this model, we see that zonal flows may affect the crossphase and hence the particle flux, via Eq. (\ref{ph-0d}).
% PARAGRAPH TO CORRECT
There are two possibilities: i) If high $\qx \fz$ gives an increase in crossphase $\cpnp$, it means it may enhance the particle flux $\Gamma$, contrary to the usual zonal flow shearing paradigm, or ii) if high $\qx \fz$ reduces the crossphase $\cpnp$, then the effect of zonal flows on crossphase simply reinforces the stabilizing effect of zonal flows on the turbulence amplitude ($\fp$), suppressing the particle flux consistent with the zonal flow shearing paradigm.
% PARAGRAPH TO CORRECT

% SIMPLIFIED MODEL & ANALYSIS

%Slaving the sideband amplitude $\fone$ to zonal flows: $\fone \sim \Lambda \fp \fz / \gamma_d$,
%We obtain, for the crossphase dynamics, at lowest order in $(\ky / \col) \ll 1$:
%\begin{eqnarray}
%\frac{\dif \cpnp}{\dif t} & \simeq & (1+\alpha \etae )\kp - \wp - \col \cpnp
%+ \Lambda  \left[ \frac{\fz \none}{\fp} ( \cpnp - \delta_1) - \frac{\nz \fone}{\fp} \cpnp \right]
 %- \col \tan \cpnp 
%\qquad \label{Kuramoto11} \\
%\frac{\dif \delta_1}{\dif t} & \simeq & (1+\alpha \etae)\kp - \omega_1  - \col \delta_1
%- \frac{\Lambda}{2} \left[ \frac{\fz \fp}{\none} (\delta_1 - \cpnp) - \frac{\nz \fp}{\none} \delta_1 \right]
%\qquad \label{Kuramoto12}
%\end{eqnarray}

%A simple physics interpretation is that the phase mismatch $\Delta \delta = \cpnp - \delta_1$ can suppress the (pump) crossphase via nonlinear-shift of the frequencies - affecting both the mode frequency $\wp$ and the normalized diamagnetic frequency $\kp$ - in the Kuramoto-like equation (\ref{Kuramoto11}). Note that a similar equation was derived in Ref. \cite{GuoDiamond2015}, without the zonal flow contribution. The associated \emph{phase-response curve} is shown [Fig. \ref{fig-phresp-hw}].
A simple physics interpretation of the zonal flow effect on crossphase is that the phase mismatch $\Delta \delta = \cpnp - \delta_1$ can suppress the (pump) crossphase via nonlinear-shift of the crossphase in the Kuramoto-like equation (\ref{ph-0d}). Note that a similar equation was derived in Ref. \cite{GuoDiamond2015}, without the zonal flow contribution, for the case of peeling-ballooning modes.
Note that Eq. (\ref{mismatch-0d}), describing the dynamics of the phase mismatch is similar to Eq. (15) of Ref. \cite{ChenLinWhite2000}, when the pump crossphase $\cpnp$ is fixed.

\begin{figure}
\begin{center}
\includegraphics[width=0.5\linewidth]{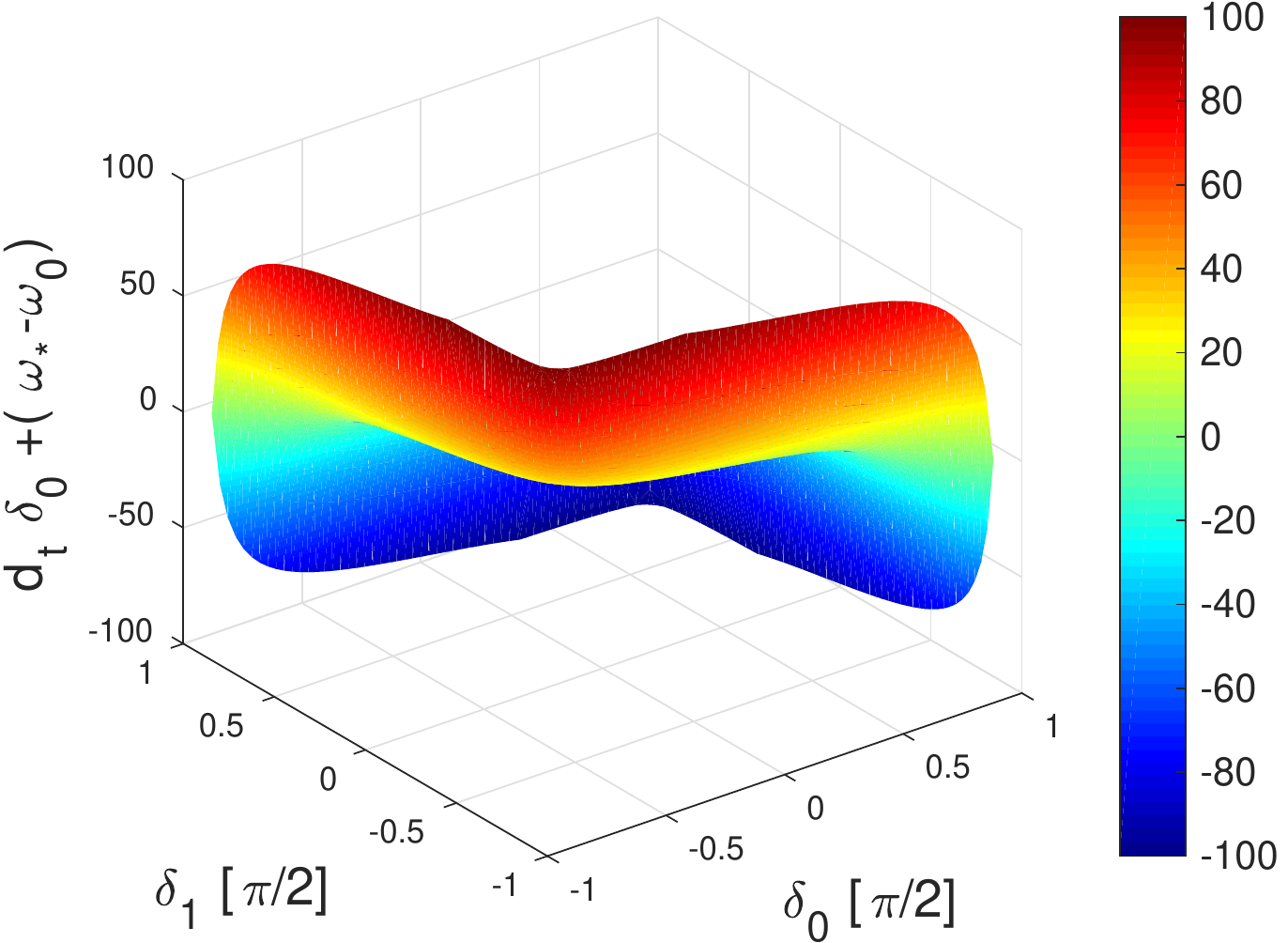}
\caption{Nonlinear phase-response curve, obtained from Eq. (\ref{ph-0dsimp}) for the DTEM model, for a normalized ZF amplitude $ \Lambda^2 \fz^2/(\col \gamma_d) = 0.5$.}
\label{fig-phresp-hw}
\end{center}
\end{figure}

%Since the solution without zonal flows is $\delta_0 \sim -\tan^{-1} [\frac{(1+ \alpha \etae)\kp- \wp}{\col}]$, and $(\kp / \col) \ll 1$, we can use the small crossphase approximation $|\cpnp|, |\delta_1| \ll 1$, and approximate Eqs. (\ref{Kuramoto11}, \ref{Kuramoto12}) as:
%\begin{eqnarray}
%\frac{\dif \cpnp}{\dif t} & \simeq & (1+\alpha \etae)\kp - \wp
 %- \col \cpnp
%+ 2 \Lambda  \left[ \frac{\fz \none}{\fp} ( \cpnp - \delta_1) - \frac{\nz \fone}{\fp} \cpnp \right]
%\label{Kuramoto21} \\
%\frac{\dif \delta_1}{\dif t} & \simeq & (1+\alpha \etae)\kp - \omega_1
%- \col \delta_1
%- \Lambda \left[ \frac{\fz \fp}{\none} (\delta_1 - \cpnp) - \frac{\nz \fp}{\none} \delta_1 \right]
%\label{Kuramoto22}
%\end{eqnarray}

%together with:
%\begin{eqnarray}
%\fone & \simeq & \frac{\kp^2 - \qx^2}{k_1^2} \frac{\Lambda}{\gamma_d} \fz \fp
%\label{sidb-0d} \\
%\none & \simeq & \frac{\Lambda}{2 \gamma_d} \Big[ \nz \fp - \fz \fp \Big]
%\label{sidbn-0d}
%\end{eqnarray}

%where ${\rm NL} (\fp, \fz)$ denotes the standard predator-prey back-reaction of zonal flows on the pump amplitude, given by:
%\begin{equation}
%{\rm NL} (\fp, \fz) = - \adw \fz^2 \fp,
%\end{equation}
%The parameter $\sigma=0$ or $1$,  is used to switch off the standard predator-prey back-reaction of zonal flows on the pump amplitude, and
%The term $- \gnl \fp^3$ is a self-damping that models same-scale interactions
%Here, $\Lambda = \qx \kp$. % and $\dia = |\nabla n| / |\nabla n|_{\rm ref}$. % and $\Lambda' = \kp^2 \Lambda$.
%The parameter $\adw = 2 \Lambda^2 / \gamma_d$ is due to DTEM interaction with zonal flows.

The evolution of total energy can be obtained by combining Eqs. (\ref{pump-0d}), (\ref{zonal-0d}), (\ref{zonaln-0d}), (\ref{sidb-0d}) and (\ref{sidbn-0d}) and is: 
\begin{align}
\frac{\dif (E_0+E_z + 2 E_1)}{\dif t} =
\trap (1+ \alpha \etae) \kp \fp^2 \cpnp - 2\trap \col (\none - \fone)^2 - \mu \qx^2 \fz^2 \notag\\
+ 2 \qx^2 \Lambda \fz \fp \fone - 2 \kp^2 \Lambda \fp \fone \fz
+ 2(\kp^2 - \qx^2) \Lambda \fone \fp \fz \notag\\
+ \trap \Lambda \nz (\fp \none - \fp \fone)
- \trap \Lambda \fp (\fz \none - \nz \fone)
- \trap \Lambda \none ( \fp \nz -  \fp \fz ) \notag\\
= \trap (1+ \alpha \etae) \kp \fp^2 \cpnp - 2\trap \col (\none - \fone)^2 - \mu \qx^2 \fz^2,
\end{align}
with $E_0 = (1+ \kp^2) \fp^2 / 2$, $E_1 = \trap \none^2 / 2 + (1+ k_1^2 - \trap) \fone^2 / 2$ and $E_z = \trap \nz^2/2 + \qx^2 \fz^2/2$.
This can be written in the form:
\begin{equation}
\frac{\dif W}{\dif t} = 2 \gamma^{nl}_{4W} W
\label{fluctdiss0}
\end{equation}
where $W=E_0+E_z+2 E_1$ the total energy, and $\gamma^{nl}_{4W}$ is the energy input-rate of the four-wave system, given by:
\begin{equation}
\gamma^{nl}_{4W} = \frac{ \trap (1+ \alpha \etae) \kp \fp^2 \cpnp - 2\trap \col (\none - \fone)^2 - \mu \qx^2 \fz^2 }{ (1+ \kp^2) \fp^2 + (1+ k_1^2 - \trap) \fone^2
+\trap \none^2  +\qx^2 \fz^2 + \trap \nz^2}
\end{equation}
This expression can be compared to Eq. (10) in Ref. \cite{Baver2002} (see also \cite{TerryGato2006}), and we recover this result for $|\cpnp| \ll 1$ and if zonal flows are neglected $\fz \to 0$.
Eq. (\ref{fluctdiss0}) can be re-expressed as:
\begin{equation}
\Gamma_{turb} = \frac{1}{1+ \alpha \etae} \frac{ \dif W}{\dif t} +\frac{\trap \col (\none-\fone)^2}{1+\alpha \etae} + \frac{ \mu \qx^2 \fz^2 }{1+ \alpha \etae},
\end{equation}
with $\Gamma_{turb} = \trap \kp \fp^2 \cpnp$ the turbulent particle flux.
This is the fluctuation-dissipation theorem for the four-wave model. It shows that, in this model, the turbulent particle flux is constrained to be non-negative, i.e. outward ($\Gamma_{turb} \ge 0$) in the saturated state ($\dif W / \dif t \sim 0$).

% SIMPLIFIED MODEL (NO ZONAL DENSITY)

\subsection*{Simplified 0D model}

Furthermore, in the limit of negligeable zonal density $\nz \ll \fz$, due to near-adiabatic response of the sideband density to potential $\none \sim \fone$ for large collisions $\col \gg \wp$, the system (\ref{ph-0d}- \ref{mismatch-0d}) reduces to:
\begin{eqnarray}
\frac{\dif \cpnp}{\dif t} & = & (1+\alpha \etae)\kp - \wp - \col \cpnp + \Lambda \frac{\fz \fone}{\fp} \phmm
\label{ph-0dsimp} \qquad\\
(1+ \kp^2) \frac{\dif \fp}{\dif t} & = & \trap (1+ \alpha \etae) \kp \fp \cpnp - 2 \kp^2 \Lambda \fone \fz
- \trap \Lambda \fone \fz
\label{pump-0dsimp} \\
\frac{\dif \fz}{\dif t} & = & 2 \Lambda \fp \fone - \mu \fz
\label{zonal-0dsimp} \\
%(1+ k_1^2) \left[ \frac{\dif \fone}{\dif t} + \gd \fone \right]  & = & (\kp^2-\qx^2 + \frac{1}{2} \trap) \Lambda \fp \fz
%\label{sidb-0dsimp} \\
\frac{\dif \phmm}{\dif t} & = & - \Delta \omega - \col \phmm - \frac{\Lambda}{2} \left( \frac{\fz \fp }{\fone} - 2 \frac{ \fz \fone}{\fp} \right) \phmm
\label{mismatch-0dsimp} %\\ 
%\frac{\dif \dia}{\dif t} & = &  - \kp \fp^2 \cpnp - \alpha N + \Gamma
%\label{profevo-0dsimp}
\end{eqnarray}
together with:
\begin{equation}
\fone \simeq \frac{ (\kp^2-\qx^2 + \frac{1}{2} \trap)}{1+k_1^2} \frac{\Lambda \fp \fz}{\gd}
\end{equation}
where $\gd$ is the sideband damping rate \cite{ChenLinWhite2000}. The \emph{phase-response curve} associated to the Kuramoto-like equation (\ref{ph-0dsimp}) is shown [Fig. \ref{fig-phresp-hw}]. This figure shows that the value of the crossphase corresponding to phase-locking condition $\dif_t \cpnp = 0$ can be shifted from its linear value $ \cpnp^{\rm lin} = [(1+ \alpha \etae) \kp - \wp] / \col$, due to the phase mismatch $\phmm = \cpnp -\delta_1$ arising from the $E \times B$ nonlinearity, last term on the r.h.s. of Eq. (\ref{ph-0dsimp}).
The dynamics of this predator-prey model Eqs.(\ref{ph-0dsimp}-\ref{mismatch-0dsimp}) is shown in the case with and without phase mismatch [Fig. \ref{fig-ppdyn}]. The associated limit-cycle is also shown in dynamical phase-space [Fig. \ref{fig-limc}].
Figure \ref{fig-ppdyn} shows the evolution of the crossphase $\cpnp$ (blue) and amplitude $\fp$ (red) associated to the pump, the zonal flow amplitude $\fz$ (yellow) and the phase mismatch $\phmm$ (magenta). The pump crossphase initially increases, driven by the density and temperature gradients and reaches the phase-locking state corresponding to the linear value $\cpnp = \cpnp^{lin}$. Then, limit-cycle oscillations occur. We now describe the typical dynamics during one cycle. The pump amplitude starts to grow exponentially, until it drives the zonal flow. As the zonal flow is driven, it back-reacts on the pump amplitude, as in the standard DW-ZF predator-prey model. Moreover, the zonal flow also \emph{back-reacts on the pump crossphase}, a novel feature of the present model. The zonal flow induces a sudden suppression of the crossphase, and the crossphase relaxes back towards its linear value, as the zonal flow amplitude decreases. Then, this cycle repeats. Without phase mismatch effects [Fig. \ref{fig-ppdyn}b], there is no
feedback on the crossphase, and thus, after phase-locking, the crossphase stays constant at its linear value, while turbulence amplitude and zonal flow amplitude undergo limit cycle oscillations.
In Fig. \ref{fig-ppdyn}, the dynamics of the model is shown in the dynamical phase-space $\fp, \fz, \cpnp - \delta^{\rm lin}$  in the case with phase mismatch (blue), and in the reference case without phase-mismatch (black). The limit-cycle is clearly visible in both cases. In the case without phase mismatch, it is two-dimensional $(\fp,\fz)$, and has a structure similar to the limit-cycle of the standard drift wave - zonal flow predator-prey model, with the growth of turbulence amplitude preceding that of zonal flow amplitude. However, when taking into account phase mismatch $\phmm \neq 0$, the limit-cycle develops a three-dimensional structure, with the transport crossphase $\cpnp$ also participating in the limit-cycle.
\begin{figure}
\begin{center}
\includegraphics[width=0.5\linewidth]{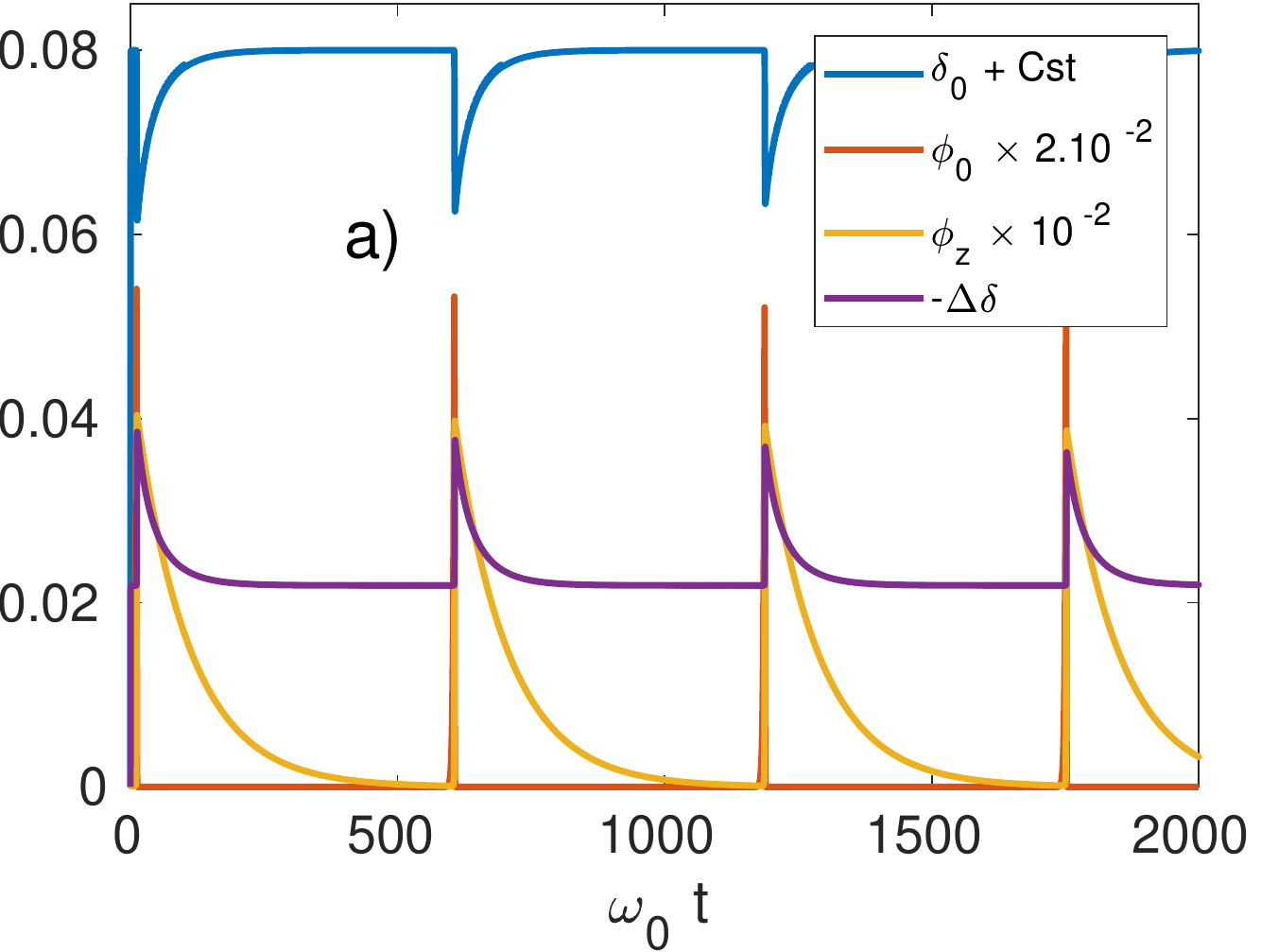}\includegraphics[width=0.5\linewidth]{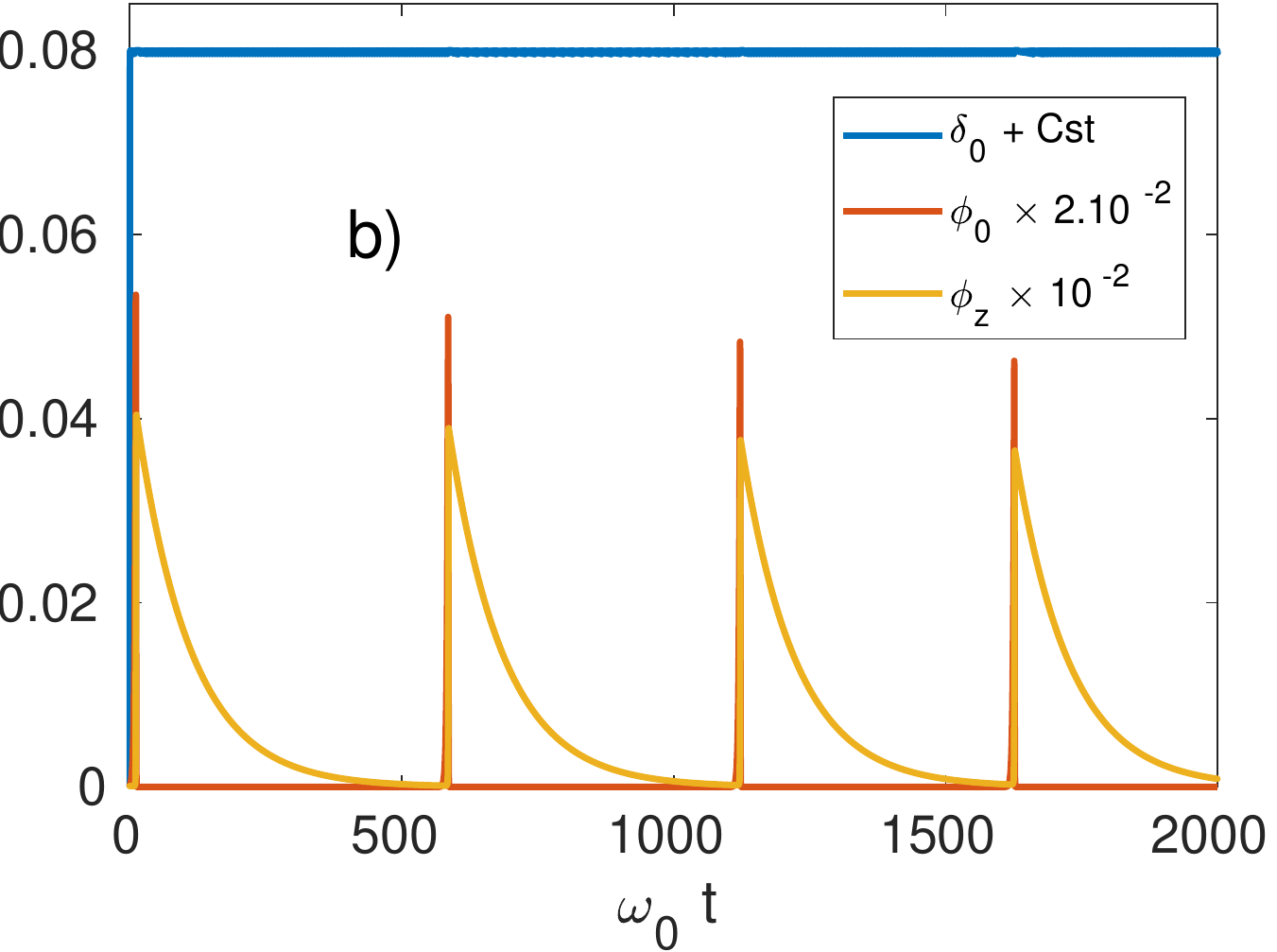}
\caption{Dynamics of the model Eqs.(\ref{ph-0dsimp}-\ref{mismatch-0dsimp}). a) with phase mismatch effects ($\phmm \neq 0$) and b) without phase mismatch effects ($\phmm = 0$) .
The parameters are: $\col =10$, $\etae=1$, $\gamma_d=0.5$, $\mu=0.01$, $\trap=0.5$. The pump and zonal wavenumbers are respectively $\kp=1$ and $\qx=0.8$.}
\label{fig-ppdyn}
\end{center}
\end{figure}

\begin{figure}
\begin{center}
\includegraphics[width=0.6\linewidth]{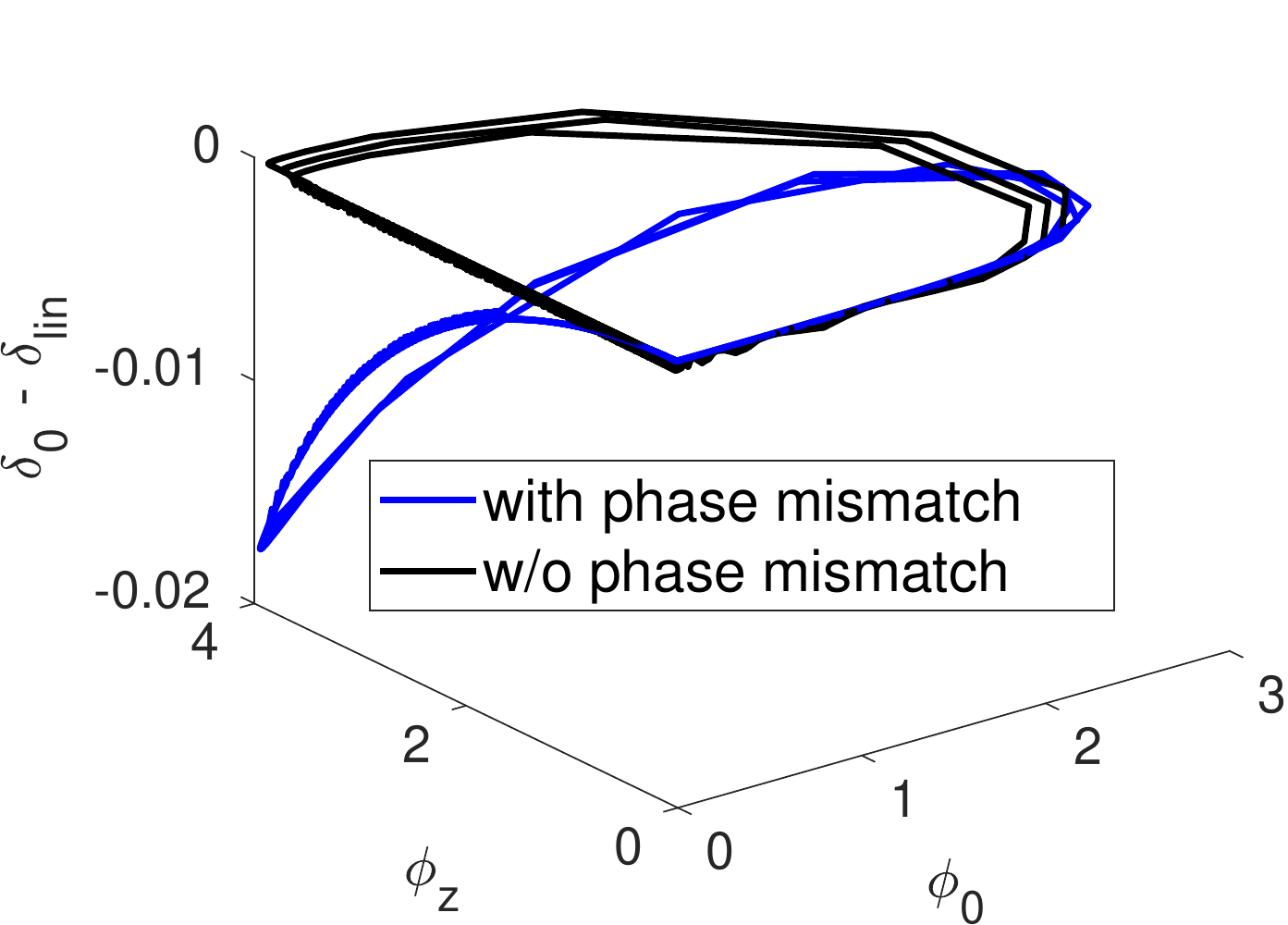}
\caption{Limit-cycle of the model with (blue) and without (black) phase mismatch effects .
The parameters are the same as in Fig. \ref{fig-ppdyn}.
}
\label{fig-limc}
\end{center}
\end{figure}

\section{Discussion and conclusions}
First, let us discuss our results concerning Eq. (\ref{cpden}) which sets the crossphase $\delta_k = \arg (n_k / \phi_k)$ between density and potential and Eq. (\ref{ampratio1}) which sets the amplitude ratio $\fk / \nk$ between density and potential.
Since the DTEM model that we use is mathematically similar (but physically very distinct) to the Hasegawa-Wakatani model describing resistive drift-waves, we compare our results to that of Ref. \cite{An2017}. However, it must be stressed that in their model, zonal modes are unphysically damped by the linear parallel dynamics (governed by the parameter named $\alpha$ in their model), whereas in the DTEM model that we use, zonal flows are not affected by the linear dynamics (the de-trapping rate $\col$ does not directly affect zonal modes). With this in mind, we can compare the crossphase equation Eq. (\ref{cpden}) with Eq. (5) of Ref. \cite{An2017}. Our analysis is slightly different, as we set the phase-angle of potential to zero, without loss of generality. This is made possible, because phase-angles are only defined \emph{up to an arbitrary phase}. Instead, the analysis in Ref. \cite{An2017} does not make use of this feature, and derives equations for both phase-angles Eqs. (3) and (4), and then substract the two equations to obtain the crossphase evolution (5). Due to this difference, our Eq. (\ref{cpden}) differs from their Eq. (5), when making the replacement $\col \to \alpha$. The first term on the r.h.s. of both Eqs. is the entrainment frequency. In Ref. \cite{An2017} it is directly the normalized electron diamagnetic frequency ($\ky$), whereas in our analysis, it is the \emph{difference} between the diamagnetic frequency $(1+ (3/2) \etae) \ky$ (including $\nabla T_e$ effects) and the mode frequency ($\omega_k$). The discrepancy is easily traced to the fact that since we set the phase of potential to zero, we need to take into account that the mode rotates at the mode frequency.
The second term on the r.h.s. of both equations is the pinning term, in our Eq. (\ref{cpden}) only the density equation contributes to this term, whereas in Eq. (5) of Ref. \cite{An2017} both the density and potential equation contribute, the reason for their $1/(k^2 \beta_k^2)$ term in the bracket of Eq. (5). The third term on the r.h.s., the nonlinear term, also  differs in their analysis compared to ours, because they include the polarization nonlinearity in the crossphase dynamics, in addition to the $E \times B$ nonlinearity.
In this respect, our analysis is more transparent, as we easily recover the results of linear analysis $\delta_k^{lin} \propto [(1+ 3/2 \etae)\ky - \omega_k] / \col$, and the amplitude ratio $\beta_k$ is self-consistently determined, Eq. (\ref{ampratio1}). We stress that our approach can be readily extended to other transport channels. For example, as we derived Eq. (\ref{cpden}) for the crossphase between potential and density using only the equation for trapped electron density, we could derive an equation for the crossphase between potential and electron temperature from an equation for trapped electron temperature. This is left for future work.
It is straightforward to extend the crossphase equation (\ref{cpden}) to include equilibrium $E \times B$ flow. The result is that \emph{both} equilibrium $E \times B$ flow or flow shear cannot \emph{directly} affect the crossphase.
This is because an equilibrium $E \times B$ flow would enter only through the Doppler-shifted frequency $\omega_k - \omega_E$ due to Galilean invariance. Since, in this case, the frequency is $\omega_k = \omega_* / (1+ k_\perp^2) + \omega_E$, the quantity $\omega_k - \omega_E$ is simply the drift-wave frequency $\omega_* / (1+ k_\perp^2)$. Hence, the effect of the equilibrium $E \times B$ flow cancels out. This is simply because fluctuations of both density and potential are equally advected by the equilibrium flow, thus the crossphase cannot change by this mechanism. However, equilibrium flow or flow shear could affect the crossphase indirectly, via its effect on zonal flows. For example, Ref. \cite{Pedrosa2008} showed experimentally that an equilibrium radial electric field - induced  by biasing the plasma - can amplify zonal flows.
On this matter, we seem to reach different conclusions than Ref. \cite{GuoDiamond2015}, where equilibrium $E \times B$ flow is claimed to directly affect the crossphase. The discrepancy may be resolved if, in Ref. \cite{GuoDiamond2015}, the $E \times B$ flow is implicitely assumed to be nonlinear (e.g. a coherent zonal flow), as we showed that zonal flows \emph{do} affect the crossphase. In this case, however, the direction of the flow is irrelevant,  as the stabilization is due to the zonal flow amplitude squared, as we showed from our 0D analysis, i.e. Eq. (\ref{ph-0dsimp}).
\quad\\
Now, we discuss the zonal density generation Eq. (\ref{zonaln-0d}) in the 0D model (\ref{ph-0d}-\ref{mismatch-0d}). In Ref. \cite{LangParkerChen2008}, it was shown that collisionless trapped electron mode turbulence  (CTEM) can saturate via nonlinear drive of zonal density. Although it is not directly relevant to our work, since we consider DTEM, we can compare the zonal density generation mechanism that we propose Eq. (\ref{zonaln-0d}) to the mechanism described by Eq. (8) in Ref. \cite{LangParkerChen2008}, as this should be model-independent. Our analysis Eq. (\ref{zonaln-0d}) differs from that of Ref. \cite{LangParkerChen2008}, in that we show that zonal density can only be driven by coupling of a pump mode $(\np, \fp)$ \emph{and} a sideband perturbation ($\none, \fone$), whereas Ref. \cite{LangParkerChen2008} claims that the pump mode can couple to itself to drive the zonal density, something that is forbidden in wave-wave interactions. It is well-known that the analysis in Ref. \cite{LangParkerChen2008} is only valid in the early growth phase of zonal density, and thus, we stress that it should not be applied to determine saturation levels. The fluid model that we use could possibly be extended to the CTEM regime, where zonal density generation seems to play a crucial role, but this is beyond the scope of this article.
\quad\\
There are limitations to our model. The trapped electron fluid model \cite{Baver2002} that we use, although taking into account the electron temperature gradient drive ($\etae$),  neglects trapped electron temperature fluctuations. This is beyond the scope of this article. Although we derive fully-nonlinear equation for the crossphase (\ref{cpden}) at the begining of this work, the four-wave approximation (parametric interaction) is used to obtain later results, in particular the nonlinear 0D model (\ref{ph-0d} - \ref{mismatch-0d}). This is a convenient method for closure of the nonlinearities, but it has the disadvantage that, from the full spectrum of wavenumbers $k$, only the pump wavenumber $\kp$ and the zonal wavenumber $q_r$ are kept, together with the sideband wavenumber. Hence, in this approach, the particle transport is implicitely assumed to be due exclusively to the pump mode. Maybe a more general approach, linking the full crossphase spectrum to zonal flows, in analogy with the wave kinetic equation (WKE) \cite{MattorDiamond1994, BrethertonGarrett1969} for the power spectrum would be preferable, but this is beyond the scope of this article. Finally, how could the theory presented in this work be tested experimentally? One possible way would be a direct test of the effect of zonal flows on the crossphase: Experiments may be able to measure directly the imaginary part - i.e. quadrature component - of the triplet correlation $ \langle \tilde n \tilde \phi \tilde n \rangle$ associated to the convective $E \times B$ nonlinearity, defined in Eq. (\ref{tripletnonlin1}) and check if the magnitude of this quantity correlates well with a change in the crossphase between density and potential fluctuations. \\
In conclusion, we showed using a parametric analysis that zonal flows can have a stabilizing effect on the transport crossphase between density and potential, in the DTEM fluid model. Future work will focus on extending this analysis to the crossphase between electron temperature and potential, and identifying possible differences in the dynamics of these two crossphases.

%It seems that what is called 'zonal density' in their analysis Eq. (8), is more like an equilibrium density (with zero radial wavenumber).
%It is also possible that the difference between their analysis and ours is due to the radial boundary condition used in their analysis, to compare with their simulation.

\section*{Acknowledgments}
The authors would like to thank J.M. Kwon and Lei Qi for usefull discussions. We also thank the anonymous Referee for helpfull comments. This work was supported by R\&D Program through National Fusion Research Institute (NFRI) funded by the Ministry of Science and ICT of the Republic of Korea (No. NFRI-EN1941-5).

% APPENDIX

% RELAXATION DYNAMICS OF PARTICLE FLUX

\section*{Appendix: Link between crossphase dynamics and relaxation dynamics of the turbulent particle flux}
In Fourier space, the model takes the form:
\begin{eqnarray}
\frac{\dif n_k}{\dif t} + (1+ \alpha \etae ) i \ky \phi_k + \col (n_k - \phi_k) & = & \notag\\
 - \frac{1}{2} \sum_{k = k' + k''} (\hat z \times k') \cdot k'' (n_{k'} \phi_{k''} - \phi_{k'} n_{k''}) \\
\frac{\dif}{\dif t} [(1 - \trap) \phi_k +k_\perp^2 \phi_k] + [1- \trap (1+\alpha \etae) ] i \ky \phi_k - \trap \col (n_k - \phi_k) & = & \notag\\
\sum_{k = k' + k''} (k_\perp'^2 - k_\perp''^2)  (\hat z \times k') \cdot k'' \phi_{k'} \phi_{k''}
\end{eqnarray}

Note the following identity:
\begin{equation}
\frac{\dif \Gamma_k}{\dif t} = \trap \ky ~ {\rm Im} \Big\{ \frac{\dif}{\dif t} (n_k^* \phi_k) \Big\}
\end{equation}
with $\Gamma_k = (1/2) \trap ~ n_k v_{rk}^* +c.c.$ the particle flux at wavenumber $k$ and $v_{rk} = - i \ky \phi_k$, and $ {\rm Im} (z) = \frac{1}{2i} (z -z^*) $.

Multiplying the c.c. of density equation by $\phi_k$ yields:
\begin{equation}
\phi_k \frac{\dif n_k^*}{\dif t} - i (1+ \alpha \etae ) \ky \fk^2 + \col ( \phi_k n_k^* - \fk^2) = 
 - \phi_k \cdot \frac{1}{2} \sum_{k = k' + k''} (\hat z \times k') \cdot k'' (n_{k'}^* \phi_{k''}^* - \phi_{k'}^* n_{k''}^*) 
\end{equation}
Multiplying the potential equation by $n_k^*$ yields:
\begin{align}
(1+k_\perp^2 - \trap) n_k^* \frac{\dif \phi_k}{\dif t} + i [1- \trap (1+\alpha \etae) ] \ky n_k^* \phi_k
- \trap \col ( \nk^2 - n_k^* \phi_k) = \notag\\
n_k^* \sum_{k = k' + k''} (k_\perp'^2 - k_\perp''^2) (\hat z \times k') \cdot k'' \phi_{k'} \phi_{k''}
\end{align}

Combining the two equations yields the evolution of (complex-valued) cross-correlation:
\begin{align}
 \frac{\dif}{\dif t} ( n_k^* \phi_k) - i (1+ \alpha \etae ) \ky \fk^2 + \col ( n_k^* \phi_k - \fk^2)
 + i \frac{ [1- \trap (1+\alpha \etae) ] \ky n_k^* \phi_k }{1 + k_\perp^2 - \trap} 
- \frac{ \trap \col ( \nk^2 - n_k^* \phi_k ) }{ 1 + k_\perp^2 - \trap }
= \notag\\
 - \phi_k \cdot \frac{1}{2} \sum_{k = k' + k''} (\hat z \times k') \cdot k'' (n_{k'}^* \phi_{k''}^* - \phi_{k'}^* n_{k''}^*)
+ \frac{n_k^*}{1 + k_\perp^2 - \trap} \sum_{k = k' + k''} (k_\perp'^2 - k_\perp''^2) (\hat z \times k') \cdot k'' \phi_{k'} \phi_{k''}
\end{align}

Finally, taking the imaginary part and multiplying by $\trap \ky$ yields the relaxation \emph{dynamics of particle flux} Eq. (\ref{fluxevo0}) in main text.

%In the parametric interaction approach, with pump amplitude $\fp$ this reduces to:
%\begin{align}
%\left[ \frac{\dif }{\dif t} + \left( 1 + \frac{\trap}{1 + \kp^2 - \trap} \right) \col \right] \Gamma_0
%- \trap (1+ \alpha \etae ) \kp^2 \fp^2
%+ \frac{ [1- \trap (1+\alpha \etae) ] \kp^2}{1 + \kp^2 - \trap} \fp^2
%= \notag\\
% - \trap \kp \cdot \frac{1}{2} (\hat z \times {\bf q}) \cdot {\bf \kp} ~
%\fp \none \fz \sin \delta_1
%+ \frac{\trap \kp}{1 + \kp^2 - \trap} \kp^2 (\hat z \times {\bf q}) \cdot {\bf \kp} ~
%\fp \fone \fz \sin \cpnp
%\end{align}

% DERIVATION OF THE 0D MODEL WITH PHASE MISMATCH

\section*{Appendix: Derivation of the 0D model}

Using the symmetry of sidebands we obtain, for $|\cpnp| \ll 1$, the following system of coupled equations:
\begin{eqnarray}
\frac{\dif \cpnp}{\dif t} & = &
(1+ \alpha \etae)\kp - \wp - \col \cpnp
+ \Lambda \left[  \frac{\fz \none}{\fp} (\cpnp - \delta_1) - \frac{\nz \fone}{\fp} \cpnp \right]
\label{den0} \qquad\\
(1+\kp^2) \frac{\dif \fp}{\dif t} & = & \trap (1+ \alpha \etae) \kp \fp \cpnp - 2 \Lambda \kp^2 \fone \fz
- \trap \Lambda ( \fz \none \cos \cpone - \nz \fone )
\label{pump0} \qquad \\
\frac{\dif \fz}{\dif t} & = & 2 \Lambda \fp \fone - \mu \fz
\label{zonal0} \\
\frac{\dif \nz}{\dif t} & = & \Lambda \left[ \fp \none \cos \cpone - \fp \fone \right]
\label{zonaln0} \\
\frac{\dif \delta_1}{\dif t} & = &
(1+ \alpha \etae)\kp - \omega_1 -\col \delta_1
- \frac{\Lambda}{2} \left[ \frac{\fz \fp}{\none} (\delta_1 - \cpnp) - \frac{\nz \fp}{\none} \delta_1 \right]
\quad \label{den1} \\
(1+ k_1^2 - \trap) \frac{\dif \fone}{\dif t}& = & \trap \col (\none \cos \cpone - \fone) + (\kp^2-\qx^2) \Lambda \fp \fz
\label{sidb0} \\
\frac{\dif \none}{\dif t} & = & - \col (\none - \fone \cos \cpone) +\frac{\Lambda}{2} \Big[ \fp \nz - \fp \fz \Big]
\label{sidbn0}
\end{eqnarray}
with the coupling coefficient $\Lambda= ({\hat z} \times {\bf q}) \cdot {\bf k}_0 = \qx \kp$, and $\cos \cpone \simeq 1 - \cpone^2/2$.

In addition, the frequencies $\wp$, $\omega_1$ are:
\begin{eqnarray}
\wp & = & \kp/(1+\kp^2) \\
\omega_1 & = & \kp / (1+\kp^2 + \qx^2)
\end{eqnarray}

Substracting Eqs. (\ref{den0}) and (\ref{den1}), we obtain the dynamics of the phase-mismatch $\phmm = \cpnp - \delta_1$ as:
\begin{equation}
\frac{\dif \phmm}{\dif t} = - \Delta \omega - \col \phmm - \frac{\Lambda}{2} \left[ \left( \frac{\fz \fp }{\none} - 2 \frac{ \fz \none}{\fp} \right) \phmm
+ \frac{\nz \fp}{\none} \delta_1 + 2 \frac{\nz \fone}{\fp} \cpnp
\right] 
\label{mismatch1}
\end{equation}
where $\Delta \omega = \wp - \omega_1 \propto \qx^2 \wp > 0$ denotes the frequency mismatch. Physically, Eq. (\ref{mismatch1}) describes the dynamics of the 'triad' phase mismatch $\cpnp - \delta_1 - \delta_q$, since we made the approximation of zero phase between zonal density and zonal potential $\delta_q = 0$.

Finally, combining Eqs. (\ref{den0}), (\ref{pump0}), (\ref{zonal0}), (\ref{zonaln0}), (\ref{sidb0}), (\ref{sidbn0}) and (\ref{mismatch1}) we obtain, after some algebra, the 0D model given in the main text (\ref{ph-0d} - \ref{mismatch-0d}).

\end{document}